\documentclass[10pt,journal,compsoc]{IEEEtran}

\usepackage{amsbsy,amsmath,amssymb,epsfig,bbm,mathrsfs,multirow,amsthm}
\usepackage{algpseudocode}
\usepackage{subcaption,float}
\usepackage[hidelinks]{hyperref}

\usepackage[noabbrev]{cleveref}
\crefrangeformat{equation}{#3(#1)#4--#5(#2)#6}

\usepackage[linesnumbered,lined,boxed,ruled,commentsnumbered]{algorithm2e}
\usepackage{blkarray}

\usepackage[T1]{fontenc}
\usepackage[utf8]{inputenc}

\usepackage{pifont}

\usepackage{ragged2e}
\usepackage{blindtext}

\newtheorem{lemma}{Lemma}
\newtheorem{theorem}{\textbf{\textsc{Theorem}}}

\DeclareMathAlphabet{\mathpzc}{OT1}{pzc}{m}{it}
\DeclareMathOperator*{\argmax}{\arg\!\max}

\usepackage{color}

\ifCLASSOPTIONcompsoc
\usepackage[nocompress]{cite}
\else
\usepackage{cite}
\fi
\ifCLASSINFOpdf
\else
\fi

\usepackage{setspace}

\begin{document}
	
	\title{Energy-based Proportional Fairness in Cooperative Edge Computing}
	\author{Thai T. Vu, Nam H. Chu, Khoa T. Phan, Dinh Thai Hoang, Diep N. Nguyen, and Eryk Dutkiewicz
	
		\IEEEcompsocitemizethanks{
			\IEEEcompsocthanksitem T. T. Vu is with the School of Engineering and Mathematical Sciences, Department of Computer Science and Information Technology, La Trobe University, Melbourne, Australia (email: t.vu@latrobe.edu.au).
			\IEEEcompsocthanksitem N.~H.~Chu, D.~T.~Hoang, D.~N.~Nguyen, E.~Dutkiewicz are with the School of Electrical and Data Engineering, University of Technology Sydney, Sydney, NSW 2007, Australia (e-mail: \{namhoai.chu, hoang.dinh, diep.nguyen, eryk.dutkiewicz\}@uts.edu.au).
			\IEEEcompsocthanksitem K.~T.~Phan is with School of Engineering and Mathematical Sciences, Department of Computer Science and Information Technology, La Trobe University, Melbourne, Australia (e-mail: K.Phan@latrobe.edu.au).		
	}
	\thanks{Preliminary results of this work will be presented at the IEEE ICC Conference 2022~\cite{vu2022energy}.}}

\IEEEtitleabstractindextext{%
\begin{abstract}
\justifying

	By executing offloaded tasks from mobile users, edge computing augments mobile devices with computing/communications resources from edge nodes (ENs), thus enabling new services/applications (e.g., real-time gaming, virtual/augmented reality). However, despite being more resourceful than mobile devices, allocating ENs' computing/communications resources to a given favorable set of users (e.g., closer to edge nodes) may block other devices from their services. This is often the case for most existing task offloading and resource allocation approaches that only aim to maximize the network social welfare or minimize the total energy consumption but do not consider the computing/battery status of each mobile device. This work develops an energy-based proportionally fair task offloading and resource allocation framework for a multi-layer cooperative edge computing network to serve all user equipments (UEs) while considering both their service requirements and individual energy/battery levels. The resulting optimization involves both binary (offloading decisions) and continuous (resource allocation) variables. To tackle the NP-hard mixed integer optimization problem, we leverage the fact that the relaxed problem is convex and propose a distributed algorithm, namely the dynamic branch-and-bound Benders decomposition (DBBD). DBBD decomposes the original problem into a master problem (MP) for the offloading decisions and multiple subproblems (SPs) for resource allocation. To quickly eliminate inefficient offloading solutions, the MP is integrated with powerful Benders cuts exploiting the ENs' resource constraints. We then develop a dynamic branch-and-bound algorithm (DBB) to efficiently solve the MP considering the load balance among ENs. The SPs can either be solved for their closed-form solutions or be solved in parallel at ENs, thus reducing the complexity. The numerical results show that the DBBD returns the optimal solution in maximizing the proportional fairness among UEs. The DBBD has higher fairness indexes, i.e., Jain's index and min-max ratio, in comparison with the existing ones that minimize the total consumed energy.	
		
\end{abstract}

}

\maketitle

\IEEEdisplaynontitleabstractindextext


\IEEEraisesectionheading{\section{Introduction}\label{sec:Int}}
	\label{sec:Introduction:TMC}
 \IEEEPARstart{S}erving an ever-growing number of mobile user equipments (UEs) calls for novel network architectures, namely edge computing~\cite{Mach2017Mobile}. In edge networks, edge nodes (ENs) are distributed closer to UEs to better serve high-demanding computing tasks, thus reducing the workload for backhaul links and enabling computation-demanding and low-latency services/applications (e.g., real-time gaming, augmented/virtual reality)~\cite{mao2017survey}. 
	However, while cloud servers, e.g., Amazon Web Services, often possess huge computing resources, an EN can provide limited computation services toward users due to its limited computing resources~\cite{el2018edge}. 
	As such, the collaboration among ENs as well as with cloud servers~\cite{du2019enabling,xing2019joint,liu2019dynamic,tran2019joint,wang2019delay,du2018enabling} to serve UEs has been considered as a very promising approach. 
	
	Moreover, offloading computing tasks from mobile devices to ENs is not always effective or even impossible due to the energy consumption for two-way data transmissions between the UEs and the ENs~\cite{Mach2017Mobile,kumar2010cloud} as well as tasks' security/QoS requirements. 
 	For that, the task offloading should be jointly optimized with the resource allocation. {\color{black}As aforementioned, despite being more resourceful than mobile devices, edge nodes' computing resource is in fact limited, especially in comparison with the cloud server. For that, allocating ENs' limited computing/communication resources to a given favorable set of users may block one or other devices from their services. This is often the case for most existing task offloading and resource allocation approaches, e.g.,~\cite{vu2021optimal,xing2019joint,wang2019delay,wang2019cooperative,wang2021fast,wang2021dependent},  that only aim to maximize the network social welfare (e.g., optimizing the total consumed energy) but not consider the computing/battery status of each mobile device. Consequently, most resources are allocated to mobile devices/services with higher marginal utilities. Whereas mobile devices with lower marginal utilities can be blocked from accessing ENs' resources. Therefore, fairness should be considered along with efficiency in edge computing. Note that the energy balancing or energy-based fairness has been also considered in IoT networks, e.g., in selecting cluster heads or routing packets to the gateway \cite{nguyen2014cooperative}. 
  
    Moreover unlike the cloud server or the edge nodes that are powered by the grid, users' mobile devices are powered by batteries hence very limited in energy. For that, it is of interest to take into account the current energy/battery levels of these mobile devices in resource allocation and task offloading decisions.   
	
	In edge computing, a few recent works consider the fairness in resource allocation and task offloading. However, all existing works are not applicable to the multi-tier edge architecture nor are able to find the optimal solution to provision the fairness among UEs given their heterogenous QoS requirements and battery levels. For example, the min-max cost policies or max-min energy balance are investigated in \cite{liu2020maxmin,du2019enabling}. 
	The authors of~\cite{du2019enabling} aim to minimize the maximum delay among mobile devices. 
	Other works, e.g., \cite{zhang2019femto,dong2019energy,li2020auction}, consider fairness amongst ENs instead of UEs.  
	For instance, the work~\cite{li2020auction} develops an auction-aided scheme that enables fair bidding for communication resources between ENs in SDN-based ultra dense networks. 
	We observe that the min-max/max-min policies only guarantee the upper bound of the cost function. They do not always provision the fairness among UEs since different UEs have different levels of resource demand. For example, it is unfair if the min-max policies are applied to two devices with $10$ and $1$ computational units, respectively. The work~\cite{liao2020blockchain} accounts for the fairness amongst user vehicles and vehicle edge servers with a heuristic reward policy. 
	Similarly, the authors of \cite{zuo2021computation} introduce a heuristic scheme for computing resource allocation that can mitigate the effect of selfish mobile users. 
	Lately, a few works investigate fairness~\cite{nguyen2018price,nguyen2019market} using market equilibrium approaches. These papers rely on the game theory and market-based frameworks, which design the price for resources in a multiple edge node and budget-constrained buyer environment. However, the market-based framework is only applicable to the two-layer model (i.e., UEs and edge node layers)~\cite{nguyen2018price,nguyen2019market}. 

 In reality, the decision to offload a task to a given edge node (out of all available edge nodes) or to the cloud must be made by considering the resource availability at the local and the edge node as well as the security requirement of the task. This is because the coupling between the task execution decision and the computing and communications resources for that purpose. Note that in any cloud/fog/edge computing platforms, all the resources must be virtualized and joint considered, e.g., the storage, the memory, the CPU/processor resource, and the networking/communications resources. Additionally, given the growing concerns on the data and users' security/privacy, different governments and industry standards have recently required cloud service providers to incorporate the security/privacy protection requirements in the service level agreement (SLA) between the service provider and the users. For example, a country's government may require all sensitive data of their citizens must be stored on servers that are within the country's physical border. Under these regulations, where a task or user data is stored or executed must take into account specific requirements of the tasks. 
	
	Given the above, this work develops an energy-based proportional-fair framework to serve all UEs with multiple tasks while considering both their service requirements (e.g., latency, security) and individual energy/battery levels in a multi-layer edge network architecture. Each UE, which may have multiple computing tasks, can connect to multiple nearby ENs to offload their tasks. The ENs can forward the tasks to a cloud server if they do not have sufficient resources to serve UEs. The edge computing and communication resources are jointly optimized with the task offloading decisions to fairly ``share'' the energy reduction/benefits to all UEs while taking into account the individual UEs' energy/battery levels. The energy/battery level at each UE is captured via a nonnegative weight factor. 
	{Finally, the load balancing among ENs is achieved by properly selecting processors for tasks (in Section~\ref{sec:balancing_processor_selection:TMC}).} 
	To the best of our knowledge, this is the first work in the literature to address the fairness of energy benefit among users in a multi-layer edge computing system with multiple tasks.}
	
	The resulting problem for offloading tasks and allocating resources toward the tasks is a Mixed Integer Nonlinear Programming (MINLP), which is generally known to be NP-hard~\cite{Boyd2004Convex}. 
	Thus, solving the problem for its optimal solution is intractable. Consequently, most of the current researches in the literature either address small-scale problems or propose approximate algorithms to find sub-optimal solutions~\cite{du2019enabling,xing2019joint,liu2019dynamic,tran2019joint}. 
	Although the main advantage of these approaches is their low complexity in finding near-optimal solutions, there is no theoretical bound/guarantee on their solutions. Instead, this work aims to find the optimal solution of the problem with our practically low complexity approach. Specifically, we leverage the convexity of its relaxed problem to propose a distributed algorithm, i.e., the dynamic branch-and-bound Benders decomposition (DBBD). The DBBD decomposes the MINLP problem according to integer variables (offloading decisions) and real variables (resource allocations) into a master problem (MP) with integer variables and subproblems (SPs) with real variables at ENs. The SPs can be then solved iteratively and parallelly at ENs until obtaining the optimal solution that meets all requirements and constraints from both UEs and ENs. Since the SPs are convex problems, it can be solved effectively by most conventional solvers for the optimal resource allocations. To support the DBBD, we develop a dynamic branch-and-bound algorithm, namely DBB, which can effectively solve the MP considering the balance between the users' demand and available resources at ENs. Thus, the load balancing among ENs can be realized. As a result, the optimal resource allocations amongst ENs, can be found at an early iteration of the DBBD. 
	Besides, the DBB is designed so that the results from solving the MP are also reused between iterations of the DBBD, thus significantly reducing the solving time compared with the conventional branch-and-bound methods~\cite{wang2019delay,narendra1977branch}. 
	The theoretical proofs and the numerical results confirm that the DBBD can always return the optimal solution maximizing the proportional fairness of the energy benefit among UEs, measured by Jain's index and the min-max ratio~\cite{jain1984quantitative}. The major contributions of this paper are summarized as follows. 
	
	\begin{itemize}
		
		\item A joint task offloading and resource allocation optimization problem that aims to maximize the fairness of energy benefits amongst UEs {while guaranteeing the load balancing among ENs} in a multi-layer edge computing network is formulated, considering both UEs' service requirements, battery levels, and ENs' resource constraints.
		
		\item To address the resulting NP-hard MINLP problem, we develop an efficient dynamic branch-and-bound Benders decomposition (DBBD) to find the globally optimal solution. Specifically, applying the Benders decomposition approach to decouple the binary and real variables, the original problem is decomposed, respectively, into a master problem (MP) for offloading selection and subproblems (SPs) for communication and computation resource allocations. We then develop a dynamic branch-and-bound algorithm (DBB) to dynamically pair computational tasks with the most potential EN, considering the tasks' demand and available resources at ENs.
		
		\item We provide theoretical analysis to demonstrate and prove the optimality and the convergence of the proposed DBBD algorithm.
		
		\item The extensive simulations confirm that the DBBD can always return the optimal solution maximizing the proportional fairness of the energy benefit among UEs, measured by Jain's index and the min-max ratio~\cite{jain1984quantitative}. 
        {\color{black}
    The results also show the superiority of DBBD in terms of fairness compared with benchmarks, e.g., FFBD~\cite{vu2021optimal}, where the total energy consumption is minimized, also called the social welfare maximization scheme (SWM)~\cite{nguyen2018price,nguyen2019market}. Extensive simulations studying the running time/complexity of the proposed algorithms confirms their practical implementation potential.
    }
	\end{itemize}
	
	{\color{black}The rest of this paper is organized as follows. Section~\ref{sec:sysmodel:TMC} describes the system model and problem formulation. In Section~\ref{sec:solutions:TMC}, we introduce the proposed optimal solution using an variant of the Benders decomposition. The complexity and performance analysis of the solution are also presented in this Section. Section~\ref{sec:performanceevaluation:TMC} presents simulations' setup and performance analysis. {Finally, Section~\ref{sec:conclusion:TMC} summarizes the major contributions and draws conclusions of this paper.}}
\section{System Model and Problem Formulation}
\label{sec:sysmodel:TMC}

\subsection{System Model}
\label{sec:sysmodel_sub:TMC}

\begin{figure}[h]
	\centering
	\includegraphics[scale=0.40]{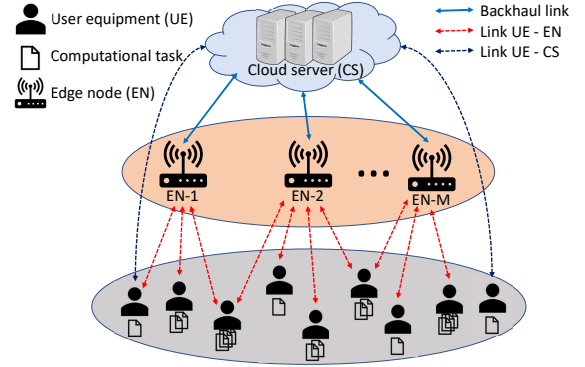}	
	\caption{Multi-layer edge computing system.}
	\label{fig:System-Model:TMC}
\end{figure}

        Consider a three-layer edge computing system in Fig.~\ref{fig:System-Model:TMC} that consists of an edge layer with $M$ edge nodes (ENs) $\mathcal{M}=\{1,\ldots,M\}$, a cloud layer with one cloud server (CS), and a user layer with $N$ user equipments (UEs) $\mathcal{N}=\{1,\ldots,N\}$. 
Let $\mathbb{Q}=\{1,\ldots,Q\}$ be the application types of computational tasks.
UEs have a set of independent computational tasks, denoted by $\Phi = \cup_{n=1}^{N}\Phi_n$, in which $\Phi_{n}$ is the set of tasks at the UE~$n$. 
{These tasks can be executed locally at UEs, offloaded to ENs, or offloaded to the CS directly from UEs or indirectly via ENs. We denote the CS in the directly offloading scenario by $V$.}
We have $\Phi_n \cap \Phi_m = \emptyset~\forall n \neq m$ and $|\Phi| = \sum_{n=1}^{N}|\Phi_n|$. Here, $|\Phi|$ be the cardinality of set $\Phi$.
{The QoS of task $I_i$ comprises the requirements of delay $t_i^r$ and security level $s_{i}^{r}$.
{\color{black}
    $t_i^r$ and $s_i^r$ are the delay/latency and the security requirements of the task $i$ (here the superscript $r$ stands for requirement to differentiate $t_i^r$ and $s_i^r$ with the actual variables/realizations of the latency $t_i$ and security $s_i$ of task $i$).
}
 {\color{black}
Our security level model was adopted from practical models (e.g., in MapReduce~\cite{dang2019trust}) and other studies that considered categorizing the security of offloaded applications into different levels~\cite{razaq2021privacy,xiao2021authentication,el2017edge}. Analogously, in our work the security requirement $s_{i}^{r}$ of application type $q$ is defined by a mapping, i.e., $s_{i}^{r}=\Theta(q) \in \mathbb{S}$ where $\mathbb{S}=\{1,\ldots, S\}$ is the security levels of UEs, ENs, and the CS. Here $1$ and $S$ respectively denote the highest and lowest levels.}
	The security requirement $s_{i}^{r}$ of application type $q$ is defined by a mapping, i.e., $s_{i}^{r}=\Theta(q) \in \mathbb{S}$. 
	Each task $I_{i}$ owned by UE $n \in \mathcal{N}$ can be then captured by $I_{i}\left(L_{i}^{u},L_{i}^{d},w_{i},t_{i}, s_{i}, q, n \right)$, in which $L_{i}^{u}$ and $L_{i}^{d}$ respectively are the input and output data size (in $\mbox{MB}$), $w_{i}$ is the number of required CPU Giga cycles per input data unit~\cite{du2019enabling}. 
	Thus, $L_{i}^{u} w_{i}$ is the required CPU Giga cycles of task $I_i$.
	Note that tasks can be processed only by UE~$n$, ENs, or the CS satisfying their QoS.}


\subsubsection{Local Processing}

The UE~$n$ has a security level $s_n\in \mathbb{S}$ and a CPU processing rate $f_{n}$. If the security requirement of task $I_i$ can be met, i.e., $s_{n} \leq s_{i}^{r}$, then task $I_i$ can be processed locally at UE~$n$. 
As in~\cite{lin2015task,chen2015decentralized,chen2016efficient}, the chip architecture of UE~$n$ can define the CPU power consumption rate as $P_{n}=\alpha(f_n)^{\gamma}$ with its specific parameters $\alpha$ and $\gamma$. Commonly, $\alpha = 10^{-11}\textnormal{Watt/cycle}^{\gamma}$ and $\gamma = 2$ since the consumed energy per operation is proportional to the square of the CPU's supply voltage, which is approximately linearly proportional to $f_n$~\cite{wen2012energy}. 
The energy consumption $E_i^l$ and the necessary computing time $T_{i}^{l}$ of UE~$n$ are given by
\begin{equation}
	E_{i}^{l}=P_{n} T_{i}^{l} = \alpha(f_n)^{\gamma} (L_{i}^{u} w_{i})/f_{n} = \alpha (f_n)^{\gamma-1} (L_{i}^{u} w_{i}) ,
	\label{eq:local_en:TMC}
\end{equation}
\begin{equation}
	T_{i}^{l}=(L_{i}^{u} w_{i})/f_{n}.
	\label{eq:local_delay:TMC}
\end{equation}

\subsubsection{Edge Node Processing}

Edge/fog node $j$ has capabilities defined by $(R_{j}^{u},R_{j}^{d},R_{j}^{f}, s_{j}^{f}, \varPsi_j)$, where $R_{j}^{u}$, $R_{j}^{d}$, $R_{j}^{f}$, $s_{j}^{f} \in \mathbb{S}$, and $\varPsi_j \subseteq \mathbb{Q}$ respectively are the total uplink, total downlink, the CPU cycle, its security level, and the set of applications supported by EN~$j$. 
If task $I_i$ is offloaded and processed at EN~$j$, then this node will allocate resources for the UE~$n$, defined by $\textbf{r}_{ij}=(r_{ij}^{u},r_{ij}^{d}, r_{ij}^{f})$, in which $r_{ij}^{u}$, $r_{ij}^{d}$ are uplink/downlink rates for transmitting the input/output, and $r_{ij}^{f}$ is the computing resource for executing the task. 
The UE~$n$ will consume an amount of energy for uploading/transmitting input data to and downloading/receiving output data from the EN~$j$. The latency of task $I_i$ comprises the time for transmission input/output and the task-execution time at EN~$j$. Let $e_{ij}^{u}$ and $e_{ij}^{d}$ be the consumed energy rates of uploading and downloading data. Let $\zeta$ be the delay caused by multi-access. 
UE~$n$ has the consumed energy $E_{ij}^{f}$ and the delay $T_{ij}^{f}$ given by
\begin{equation}
	E_{ij}^{f}=e_{ij}^{u}L_{i}^{u} +e_{ij}^{d}L_{i}^{d},
	\label{eq:edge_en:TMC}
\end{equation}
\begin{equation}
	T_{ij}^{f}=L_{i}^{u}/r_{ij}^{u}+L_{i}^{d}/r_{ij}^{d} + (L_{i}^{u} w_{i})/r_{ij}^{f} + \zeta.
	\label{eq:edge_delay:TMC}
\end{equation}
{\color{black}
    Note that since task $I_i$ is not executed at UE $n$, the total energy consumed at the UE $n$ only consists of the energy for uploading the task and downloading the data/result (but not include the energy to processing the task, that is executed by the edge node), as considered in Eq.~(\ref{eq:edge_en:TMC}).
}
{Various factors of the wireless environment like interference, channel fading can be captured by the consumed energy~$E_{ij}^{f}$ and delay~$T_{ij}^{f}$.}

\subsubsection{Cloud Server Processing (offloaded via an edge node)}


Let $\mathcal{B} = \{\mathcal{B}_{1},\ldots,\mathcal{B}_{M}\} \in \mathbb{R}^{M}$ be the backhaul capacity between $M$ ENs and the CS.
All tasks offloaded to the CS via EN~$j$ will share the backhaul $\mathcal{B}_j$. Let $\mathcal{C} = \{\mathcal{C}_{1},\ldots,\mathcal{C}_{Q}\} \in \mathbb{R}^{Q}$ be the processing rate the CS can allocate to each task of $Q$ applications.
Let $s_{q}^{c}$ be the security level of the CS toward application~$q$. If the security requirement is satisfied, i.e., $s_{i}^{r} \geq s_{q}^{c}$, then EN~$j$ can forward task $I_i$ to the CS.


In this case, the EN~$j$ will allocate resources $\mathbf{r}_{ij}=(r_{ij}^{u},r_{ij}^{d},r_{ij}^{f})$ for the UE~$n$, where $r_{ij}^{u}$, $r_{ij}^{d}$ are uplink/downlink rates for transmitting input/output data, and $r_{ij}^{f}\!=\!0$ (since it does not process the task). Then, the CS will allocate backhaul rate $b_{ij}$ to transmitting input/output data between EN~$j$ and the CS. Task~$i$ will be processed at the CS with computation rate $\mathcal{C}_{q}$. 
The energy consumption $E_{ij}^{c}$ at the UE includes  the energy for transmitting/receiving input/output to and from EN~$j$. The delay $T_{ij}^{c}$ comprises the time for transmitting the input from the UE to the CS via FN~$j$, the time for receiving the output from the CS via the EN~$j$, and the time for executing the task at the CS. These metrics are given by
\begin{equation}
	E_{ij}^{c}=E_{ij}^{f}=e_{ij}^{u}L_{i}^{u} +e_{ij}^{d}L_{i}^{d},
	\label{eq:indirect_cloud_en:TMC}
\end{equation}
\begin{equation}
	T_{ij}^{c}=L_{i}^{u}/r_{ij}^{u}+L_{i}^{d}/r_{ij}^{d}+
	(L_{i}^{u}+L_{i}^{d})/b_{ij}+(L_{i}^{u} w_{i})/\mathcal{C}_{q} + \zeta.
	\label{eq:indirect_cloud_delay:TMC}
\end{equation}

\subsubsection{Cloud Server Processing (directly offloaded by user equipment)}
\label{sub:direct_cloud_server_processing:TMC}

To simplify the notation, in the sequel we denote the cloud $V$ as an extra edge node, i.e., $(M+1)$-th~EN, of the set $\mathcal{M^*} \!=\! \mathcal{M} \cup \{V\}$.  All UEs share resources denoted by $(R_{(M+1)}^{u},R_{(M+1)}^{d},R_{(M+1)}^{f}, s_{(M+1)}^{f}, \varPsi_{(M+1)})$ for the direct connection to the cloud. Here, $R_{(M+1)}^{u}$, $R_{(M+1)}^{d}$, and $R_{(M+1)}^{f}$ are total uplink, downlink, and the CPU cycle of the cloud, respectively. The security level $s_{(M+1)}^{f}$ is defined as $s_{(M+1)}^{f}=s_{q}^{c}$ for the application type~$q$ and $\varPsi_{(M+1)} = \mathbb{Q}$ shows that the cloud $V$ can support all application types.  

If task $I_i$ is directly offloaded to the cloud~$V$ (or the $(M+1)$-th node) for execution, the cloud will allocate uplink/downlink communication and computation resources toward UE~$n$, denoted $\mathbf{r}_{i(M+1)}=$ $(r_{i(M+1)}^{u},$ $r_{i(M+1)}^{d},$ $r_{i(M+1)}^{f})$, for input/output transmission and executing the task. In this case, the consumed energy of the UE $E_{i(M+1)}^{f}$ and the delay $T_{i(M+1)}^{f}$ are similar to those in Eqs.~(\ref{eq:edge_en:TMC})~and~(\ref{eq:edge_delay:TMC}) and given by
\begin{equation}
	E_{i(M+1)}^{f}=e_{i(M+1)}^{u}L_{i}^{u}+e_{i(M+1)}^{d}L_{i}^{d},
	\label{eq:direct_cloud_en:TMC}
\end{equation}
\begin{equation}
	\label{eq:direct_cloud_delay:TMC}
	\begin{split}
		T_{i(M+1)}^{f}=&L_{i}^{u}/r_{i(M+1)}^{u}+L_{i}^{d}/r_{i(M+1)}^{d} \\
		&+ (L_{i}^{u} w_{i})/r_{i(M+1)}^{f} + \zeta,
	\end{split}
\end{equation}
where $e_{i(M+1)}^{u}$ and $e_{i(M+1)}^{d}$ are the energy consumption for directly transmitting and receiving a unit of data between the UE~$n$ and the cloud.
Since the cloud (i.e., $(M+1)$-th node) is the top layer, it cannot forward the task to a higher layer. Mathematically, this is captured by setting the backhaul capacity to the higher layer $\mathcal{B}_{(M+1)} = 0$.

\subsubsection{Task Categorization}
\label{sec:task_categorization:TMC}

If a decision to process a computational task requires less energy than the worst case (i.e., local processing or the worst case of offloading), then the energy difference (between this decision and the worst case) is referred to as the energy benefit of the decision. This work aims to maximize the fairness in terms of energy benefits amongst the devices when processing their tasks. For each task, from the energy consumed by the UE in Eqs.~(\ref{eq:local_en:TMC}),~(\ref{eq:edge_en:TMC}),~(\ref{eq:indirect_cloud_en:TMC}),~and~(\ref{eq:direct_cloud_en:TMC}) for four possible decisions and the task's corresponding QoS requirement, we classify it into one of the four categories in  Table~\ref{tab:task_categories:TMC}. 


{\renewcommand{\baselinestretch}{1} 
	\begin{table}[h]
		\caption{Categories of tasks according to energy benefits and QoS satisfaction. \label{tab:task_categories:TMC}}
		\centering{}%
		\footnotesize
		\begin{tabular}{|l|c|c|c|c|}
			
			\hline
			\textbf{Task} &
			\textbf{Local} & \textbf{Offloading} & \textbf{Offloading} &\textbf{Pre-decision} 
			\\
			\textbf{Categories} &
			\textbf{QoS} &
			\textbf{QoS} &
			\textbf{Benefits} &
			\tabularnewline
			\hline
			
			\textbf{Cat-1} & \ding{51} & -- & \ding{53}  & pre-local \tabularnewline
			\hline
			
			\textbf{Cat-2} & \ding{51} & \ding{53} & \ding{51}  & pre-local \tabularnewline
			\hline
			
			\textbf{Cat-3} ($\widehat{\Phi}$) & \ding{51} & \ding{51} & \ding{51}  & local, offload \tabularnewline	
			\hline
			
			\textbf{Cat-4} ($\widetilde{\Phi}$) & \ding{53} & \ding{51} & --  & offload \tabularnewline
			\hline
			
		\end{tabular}
	\end{table}
}

{\color{black}
For the tasks in \textbf{Cat-1} and \textbf{Cat-2}, while local execution satisfies the QoS requirements, the offloading does not lead to the energy benefits and the QoS satisfaction, respectively, thus they are predetermined to be processed locally.
For the tasks in \textbf{Cat-3}, the QoS requirements are satisfied by both local processing and offloading. Thus, depending on the available resources at ENs and the cloud, the tasks can be either processed locally or offloaded. 
For the tasks in \textbf{Cat-4}, only offloading can meet their QoS requirements, thus they are predetermined to be offloaded. Note that for a task that is demanding in both latency and security requirement, e.g., when the local processing (with the highest security level requirement) cannot meet the strict latency requirement (i.e., high computing capability is required), the QoS cannot be met and the task must be aborted. In practice, QoS standards and protocols or similar ones at the application layer, e.g., DiffServ/NPLS can help ensure the QoS requirements to be met. For that, we do not consider the tasks whose QoS cannot be met here. Classifying all tasks into these categories can be carried out at the pre-processing stage.}

Since computational tasks in \textbf{Cat-1} and \textbf{Cat-2} are predetermined to be processed locally, without loss of generality, we assume that $\Phi$ contains computational tasks only belonging to \textbf{Cat-3} and \textbf{Cat-4}, denoted $\widehat{\Phi}$ and $\widetilde{\Phi}$, respectively. Equivalently, for UE~$n$, let $\widehat{\Phi}_n$ and $\widetilde{\Phi}_n$ respectively be the sets of its tasks belonging to \textbf{Cat-3} and \textbf{Cat-4}. Thus, we have $\Phi = \widehat{\Phi} \cup \widetilde{\Phi}$ and $\Phi_n = \widehat{\Phi}_n \cup \widetilde{\Phi}_n$.



\subsection{Problem Formulation}
\label{sec:problemformulation:TMC}

The offloading decisions of task $I_{i}$ can be modelled as $\mathbf{x}_{i}=$ $(x_{i}^{l},x_{i1}^{f} \dots,x_{i(M+1)}^{f},$ $x_{i1}^{c} \dots,x_{i(M+1)}^{c})$, where either $x_{i}^{l}=1$ or $x_{ij}^{f}=1$ or $x_{ij}^{c}=1$ determines that task $I_{i}$ is exclusively executed at either the UE or EN~$j$ or the CS (via EN~$j$). Equivalently, we have the delay and energy consumption, i.e., $\mathbf{h}_i =$ $(T_i^l, T_{i1}^f,\dots,T_{i(M+1)}^f,$ $T_{i1}^c,\dots,T_{i(M+1)}^c)$ and $\mathbf{e}_i = (E_{i}^l, E_{i1}^f,\dots,E_{i(M+1)}^f,E_{i1}^c,\dots,E_{i(M+1)}^c)$, as in Eqs.~(\ref{eq:local_en:TMC})--(\ref{eq:direct_cloud_delay:TMC}). Due to either the energy, the computing limitations, or the security requirements, not all the tasks can be processed locally (e.g., local processing cannot satisfy the delay requirement). For that, we classify the set of tasks ${\Phi}$ into two categories: $\widehat{\Phi}$ for tasks that can be either executed locally or offloaded and $\widetilde{\Phi}$ for tasks that are unable to be executed locally and always need to be offloaded. 

Let $E_{i}^{base}$ denote the total energy consumption required for the \emph{baseline solution} to execute task $I_i$, depending on which category the task belongs to. 
Thus, we can define $E_{i}^{base}$ as
\begin{equation}
	E_{i}^{base}=\begin{cases}
		E_{i}^{l}, I_i \in \widehat{\Phi}, \\
		\underset{j \leq (M+1) }{\max}\{E_{ij}^f,E_{ij}^c\} \!=\! \underset{j \leq (M+1) }{\max}\{E_{ij}^f\},I_i \in \widetilde{\Phi}.
	\end{cases} \label{eq:e_base:TMC}
\end{equation}
Then, the energy benefit/saving $\Delta_{i}$ of UE 
in comparison with the base line $E_{i}^{base}$ 
and the task-execution delay $T_i$ are given by
\begin{equation}
	\Delta_{i}= (E_{i}^{base} - \mathbf{e}_i)^{\top}\mathbf{x}_{i},
	\label{eq:task_en:TMC}
\end{equation}
\begin{equation}
	T_{i}=\mathbf{h}_i^{\top}\mathbf{x}_{i} .
	\label{eq:task_delay_nonconvex:TMC}
\end{equation}
The relaxation of $T_{i}$ in Eq.~(\ref{eq:task_delay_nonconvex:TMC}) is not convex due to its factors of the form $x/r$ where $x$ and $r$ are the offloading decision and resource allocation variables.  
Consequently, the optimization problem with the delay constraints, i.e., $T_{i} \leq t_i^r$, is not convex. 
To leverage convexity in finding the optimal solution, we transform $T_i$ in Eq.~(\ref{eq:task_delay_nonconvex:TMC}) to an equivalent convex one. Let  $\mathbf{y}_{i}=((x_{i}^{l})^2,(x_{i1}^{f})^2, \dots,(x_{i(M+1)}^{f})^2,(x_{i1}^{c})^2, \dots,(x_{i(M+1)}^{c})^2)$. We have $\mathbf{y}_{i} = \mathbf{x}_{i}$ due to its binary variables $x_{i}^{l}$, $x_{ij}^{f}$, and $x_{ij}^{c}$. 
In the remainder of this paper, the delay $T_i$ in Eq.~(\ref{eq:task_delay_convex:TMC}) will be used for task $I_i$. We will prove its convexity in Theorem~\ref{theo:convexity:TMC}.
\begin{equation}
	T_{i}=\mathbf{h}_i^{\top}\mathbf{y}_{i}.
	\label{eq:task_delay_convex:TMC}
\end{equation}
	Finally, we can define the utility function of the UE~$n$ with its set of tasks $\Phi_n$ as follows.
	\begin{equation}
		\label{eq:utility_fuc_:TMC}
		u_n = \sum_{I_i \in \Phi_n}\Delta_{i}.
	\end{equation}
	
	
	Without considering the fairness in energy reduction/benefit for users in the task offloading decision, one can simply optimize the total of all individual users' utility functions $u_n$ in Eq.~(\ref{eq:utility_fuc_:TMC}). As a result, the UEs owning the tasks with less energy benefit $\Delta_{i}$ may not be served by any EN. Consequently, these UEs will soon run out of energy and fail to maintain their functions. 
	Unlike existing works, e.g., \cite{xing2019joint,wang2019delay,wang2019cooperative,wang2021fast,wang2021dependent}, this paper addresses the problem of joint task-offloading (\textbf{x}) and resource-allocating $(\mathbf{r}, \mathbf{b})=(\{\mathbf{r}_{ij} \},\{{b}_{ij} \} )$ 
	so that all UEs can achieve their proportionally fair share of energy benefit/saving, considering their delay, security, application compatibility requirements as well as their battery/energy status.
	Let $\rho_n\in[0,1]$ be the weight of the UE~$n$ that captures the user's battery status/priority level. 
	Without loss of generality, we assume that $\forall n, \Phi_n$ is not empty and always has a task $I_i$ with positive energy benefit, i.e.,  $\Delta_{i} > 0$. In other words, $u_n > 0, \forall n \in \mathcal{N}$.
	
	As defined in \cite{kelly1998rate}, a vector of utility functions \mbox{$\textbf{u}=(u_1, \dots, u_N)$} of $N$ UEs is {proportionally fair}, if it is feasible, i.e., there exists an offloading and resource allocation solution satisfying  $\textbf{u} \succ \textbf{0}$, and meeting all requirements and constraints from both UEs and ENs. Here, $\succ$ denotes componentwise inequality. In addition, for any other feasible vector $\textbf{u}^{*}$ regarding the proportional fairness over the weight $\rho_n$ of each UE~$n$, the aggregation of proportional changes from $\textbf{u}$ is not positive as~\cite{kelly1998rate}
	\begin{equation}
		\label{eq:fairness_def:TMC}
		\sum_{n=1}^{N} \rho_n \frac{u_n^{*} - u_n}{u_n} \leq 0.
	\end{equation}
	In other words, the total of proportional benefit changes of any solution $\textbf{u}^{*}$ comparing with $\textbf{u}$ is less or equal to $0$.
	Equivalently, Eq.~(\ref{eq:fairness_def:TMC}) can be rewritten in the derivative form as follows.
	\begin{equation}
		\label{eq:fairness_def_log:TMC}
		\sum_{n=1}^{N} \rho_n (\text{ln}(u_n))^{'}du_n \leq 0.
	\end{equation}

	From Eq.~(\ref{eq:fairness_def_log:TMC}), 
	the proportionally fair joint offloading and resource allocation solution can be obtained by maximizing of the utility function $\sum_{n=1}^{N} \rho_n \text{ln}(u_n)$ over offloading decision $(\mathbf{x})$ and resource allocation variables $(\mathbf{r}=\{\mathbf{r}_{ij} \}~\textnormal{and}~ \mathbf{b}=\{{b}_{ij} \} )$ for all tasks in $\Phi = \widetilde{\Phi} \cup \widehat{\Phi}$. 
	The equivalent optimization problem considering tasks' QoS requirements and edge nodes' resource constraints is formulated as follows.
	\begin{equation}
		(\mathbf{P}_0) \hphantom{5}	 \underset{\mathbf{x},\mathbf{r},\mathbf{b}}{\max} ~\sum_{n=1}^{N} \rho_n \text{ln}(u_n) , ~\text{s.t.}~(\mathbf{R}_0)~\text{and}~(\mathbf{X}_0),
		\label{eq:global_en_merge:TMC}
	\end{equation}
	\begin{equation}
		\begin{split}
			(\mathbf{R}_0)
			\begin{aligned}
				\left\{	\begin{array}{ll}
					(\mathcal{C}_1) \hphantom{1} T_{i} \leq t_{i}^{r}, \forall i\in \Phi,	\\
					(\mathcal{C}_2)  \hphantom{1} \sum \limits_{i \in \Phi}r_{ij}^{f} \leq R_{j}^{f}, \forall j\in \mathcal{M^*},	\\
					(\mathcal{C}_3)  \hphantom{1} \sum \limits_{i \in \Phi}r_{ij}^{u} \leq R_{j}^{u}, \forall j\in \mathcal{M^*},	\\
					(\mathcal{C}_4)  \hphantom{1} \sum \limits_{i \in \Phi}r_{ij}^{d} \leq R_{j}^{d}, \forall j\in \mathcal{M^*},	\\
					(\mathcal{C}_5)  \hphantom{1} \sum \limits_{i \in \Phi}b_{ij} \leq \mathcal{B}_j, \forall j\in \mathcal{M^*},	\\
					r_{ij}^{u}, r_{ij}^{d}, r_{ij}^{f}, b_{ij} \geq 0,\forall (i,j)\in \Phi\times\mathcal{M^*},
				\end{array}	\right.
			\end{aligned} 
			\\ 
			(\mathbf{X}_0)	
			\begin{aligned}
				\left\{	\begin{array}{ll}				
					(\mathcal{C}_6) \hphantom{0} x_{i}^{l}+\sum \limits_{j=1}^{M+1}x_{ij}^{f}+\sum \limits_{{j=1}}^{M+1}x_{{ij}}^{{c}}=1, \forall i\in \Phi,	\\				
					(\mathcal{C}_7) \hphantom{0} x_{i}^{l} s_{i}^{l} + \sum \limits_{{j=1}}^{M+1} x_{{ij}}^{{f}} s_{j}^{f} + \sum \limits_{j=1}^{M+1} x_{{ij}}^{{c}} s_{q}^{c} \leq s_{i}^{r}, \\ \forall i\in \Phi, \\				
					(\mathcal{C}_8) \hphantom{0} x_{ij}^{f} = 0, \forall (i,j) \in \Phi \times \overline{\mathbb{G}}(q), \\				
					(\mathcal{C}_9) \hphantom{1} x_{i}^{l} = 0, \forall i \in \widetilde{\Phi}, \\				
					x_{i}^{l},x_{ij}^{f}, x_{ij}^{c} \in\{0,1\}, \forall (i,j)\in \Phi\times\mathcal{M^*},				
				\end{array}	\right.
			\end{aligned}
		\end{split}
		\label{eq:resources_and_offload_con_merge:TMC}
	\end{equation}
	where $(\mathcal{C}_1)$, $(\mathcal{C}_7)$, and $(\mathcal{C}_8)$ capture tasks' QoS requirements, i.e., the delay, security, and application compatibility; $(\mathcal{C}_2)$, $(\mathcal{C}_3)$, $(\mathcal{C}_4)$, and $(\mathcal{C}_5)$ capture ENs' resource bounds, i.e., the computational, uplink, downlink, and backhaul; $(\mathcal{C}_6)$ guarantees that a task is exclusively processed locally, at an EN, or at the cloud;  $(\mathcal{C}_9)$ specifies tasks that can't be processed locally; $\overline{\mathbb{G}}(q)$ is the set of all ENs that do not support the application type $q$. As defined in Section~\ref{sub:direct_cloud_server_processing:TMC}, $\mathcal{M^*} = \mathcal{M} \cup \{V\}$.

{\color{black} {\bf{Remark:}} Under the theoretical framework laid by F. Kelly~\cite{kelly1998rate}, the individual utility maximization problem, i.e., maximizing the $u_n$ equation above, involving the binary variables is hence not convex. For that, the optimal solution of $(\mathbf{P}_0)$ does not theoretically guarantee the proportional fairness. However, like other seminal works in the literature, e.g., ones that employed the Nash Bargaining framework, \cite{han2005fair,nguyen2015distributed} (a generalized version of the problem considered by F. Kelly), the fairness of our proposed framework above is empirically verified in the experiment section.}
	
	\section{Proposed Optimal Solutions}
	\label{sec:solutions:TMC}
	
	As aforementioned, the $(\mathbf{P}_0)$ is NP-hard due to its binary ($\mathbf{x}$) and real variables ($\mathbf{r},\mathbf{b}$). In general, it is intractable to find its optimal solution. However, by relaxing the integer variables to real numbers, the resulting relaxation of $(\mathbf{P}_0)$ becomes a convex optimization problem~\cite{Boyd2004Convex}. In the sequel, we leverage this feature to develop an effective algorithm to find the optimal solution of $(\mathbf{P}_0)$.
	
	\subsection {Convexity of Relaxed Problem}
	
	From the original problem $(\mathbf{P}_0)$, we can transform it into a fully-relaxed problem as follows.
	\begin{equation}
		(\mathbf{\widetilde{P}}_0) \hphantom{10}	 \underset{\mathbf{x},\mathbf{r},\mathbf{b}}{\max} ~\sum_{n=1}^{N} \rho_n \text{ln}(u_n) ,~\text{s.t.}~(\mathbf{R}_0)~\text{and}~(\mathbf{\widetilde{X}}_0),
		\label{eq:global_en_relax:TMC}
	\end{equation}
	\begin{equation}
		(\mathbf{\widetilde{X}}_0)
		\begin{aligned}
			\left\{	\begin{array}{ll}				
				(\mathcal{C}_6),~(\mathcal{C}_7),~(\mathcal{C}_8),~(\mathcal{C}_9) \\				
				x_{i}^{l},x_{ij}^{f}, x_{ij}^{c} \in[0,1], \forall (i,j)\in \Phi \!\times\!\mathcal{M^*}.				
			\end{array}	\right.
		\end{aligned}
		\label{eq:17_offload_relax_con:TMC}
	\end{equation}	
	By converting all binary variables to real numbers, i.e., $x_i^l,x_{ij}^f,x_{ij}^c \in [0,1], \forall (i,j) \in \Phi \times \mathcal{M^*}$, the problem $(\mathbf{\widetilde{P}}_0)$ is a standard nonlinear problem. Theorem~\ref{theo:convexity:TMC} below proves the convexity of $(\mathbf{\widetilde{P}}_0)$.

	\begin{theorem}
		\label{theo:convexity:TMC}
		The optimization problem $(\mathbf{\widetilde{P}}_0)$ is convex.
	\end{theorem}
	
	
	{\emph{Proof:}} The proof is shown in Appendix~\ref{sec:theo_convexity:TMC}.
	
	
	
	\subsection{Dynamic Branch-and-Bound Benders Decomposition}

	\begin{figure}[h]
		\centering
		\includegraphics[scale=0.45]{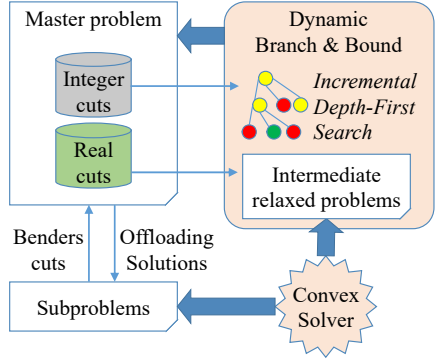}
		\caption{Dynamic branch-and-bound Benders decomposition.}
		\label{fig:DBBD_Model:TMC}
	\end{figure}
	
	We introduce a dynamic branch-and-bound Benders decomposition (DBBD), as in Fig.~\ref{fig:DBBD_Model:TMC}.  
	In DBBD, $(\mathbf{P}_0)$ is first decomposed according to integer variables (offloading decisions) and real variables (resource allocations) into a master problem $(\mathbf{MP}_0)$ with the integer variables and subproblems $(\mathbf{SP}_0)$ with the real variables. Then, we develop a dynamic branch-and-bound algorithm (DBB), which is equipped with an incremental depth-first search, to quickly find the optimal offloading solution of the $(\mathbf{MP}_0)$. 
	The \textit{Benders cuts} that eliminate inefficient solutions of $(\mathbf{SP}_0)$ are generated and updated in the $(\mathbf{MP}_0)$. The DBBD finds the optimal solution of $(\mathbf{P}_0)$ by iteratively solving $(\mathbf{MP}_0)$ and $(\mathbf{SP}_0)$. 
	\begin{equation}
		\begin{split}
			(\mathbf{MP}_0)~\mathbf{x}^{(k)}=\underset{\mathbf{x}}{\argmax} \{\sum_{n=1}^{N} \rho_n \text{ln}(u_n) \}, \text{s.t.}~ cuts^{(k)}~\text{and}~(\mathbf{X}_0),
		\end{split}		
		\label{eq:master_problem:TMC}
	\end{equation}
	\begin{flalign}		
		(\mathbf{SP}_0) \hphantom{10}	 \underset{\mathbf{r},\mathbf{b}}{\min}\{0\},~\text{s.t.}~(\mathbf{R}_0) , &&
		\label{eq:subproblem:TMC}
	\end{flalign}
	where $cuts^{(k)}$ is the set of Benders cuts generated at previous iterations $(1,\dots, (k-1))$ as in Section~\ref{sub:Cutting_Plane_Generation:TMC},   $\{0\}$ is the constant zero. Here, $cuts^{(k)}$ are constraints on offloading variables $\mathbf{x}$ of $(\mathbf{MP}_0)$ at iteration $(k)$.
	
	In DBBD, at iteration~$(k)$, $(\mathbf{MP}_0)$ first is solved to find the offloading solution, i.e.,  $\mathbf{x}^{(k)}$, of $(\mathbf{MP}_0)$. Then, the $(\mathbf{SP}_0)$ is solved to find the resource allocation, i.e., $(\mathbf{r}, \mathbf{b})$, toward the offloaded tasks that are determined by $\mathbf{x}^{(k)}$ of $(\mathbf{MP}_0)$. 
	According to Theorem~\ref{theo:dbbdstop:TMC}, DBBD can terminate the iteration if either $(\mathbf{MP}_0)$ is infeasible or it returns a solution $(\mathbf{x}, \mathbf{r}, \mathbf{b})$. 
	
	
	
	\begin{theorem}
		\label{theo:dbbdstop:TMC}
		At iteration $(k)$, if a solution $(\mathbf{x})$ of $(\mathbf{MP}_0)$ leads to a solution $(\mathbf{r},\mathbf{b})$ of $(\mathbf{SP}_0)$, then $(\mathbf{x},\mathbf{r},\mathbf{b})$ is the optimal one of $(\mathbf{P}_0)$. In addition, at iteration $(k)$, if $(\mathbf{MP}_0)$ is infeasible, then $(\mathbf{P}_0)$ is infeasible.
	\end{theorem}
	
	{\emph{Proof:}} The proof is shown in Appendix~\ref{sec:theo_dbbdstop:TMC}.
	
	
	\subsection{Distributed/parallel Subproblems}
	
	At iteration $(k)$, the offloading decision $\mathbf{x}^{(k)}$ helps to break down $(\mathbf{SP}_0)$ into $(M+1)$ smaller independent problems $(\mathbf{SP}_1)$ at $(M+1)$ ENs (including the cloud server $V$). The resource allocation problem $(\mathbf{SP}_1)$ at EN~$j$ is for tasks that are offloaded to EN~$j$ (denoted $\Phi_j^t$) and to the CS via EN~$j$ (denoted $\Phi_j^s$). Equivalently, these sets are captured by $\mathbf{x}_j^{f(k)} = (x_{1j}^{f}, \dots, x_{|\Phi|j}^{f})^{(k)}$ and $\mathbf{x}_j^{c(k)} = (x_{1j}^{c}, \dots, x_{|\Phi|j}^{c})^{(k)}$ in $\mathbf{x}^{(k)}$. Thus, we can define $\Phi_j^t = \{1,\dots t\}$, $\Phi_j^s = \{t+1, \dots t+s\}$, and $\Phi_j^{t+s} =\Phi_j^t \cup \Phi_j^s= \{1,\dots,t+s\}$ is captured by $\mathbf{x}_j^{(k)} = (\mathbf{x}_j^{f(k)},\mathbf{x}_j^{c(k)})$. Variables $\mathbf{r}_j= (\mathbf{r}_{1j}, \dots \mathbf{r}_{(t+s)j})$ and $\mathbf{b}_j= (b_{1j}, \dots b_{(t+s)j})$ denote resource allocation of EN~$j$ towards the set of tasks $\Phi_j^{t+s}$. The problem $(\mathbf{SP}_1)$ at EN~$j$ can be defined as
	\begin{equation}
		(\mathbf{SP}_1) \hphantom{10}	 
		\underset{\mathbf{r}_j, \mathbf{b}_j}{\min}\{0\} , ~\text{s.t.}~(\mathbf{R}_j),
		\label{eq:sub_constrant_goal:TMC}
	\end{equation}
	\begin{equation}
		\begin{aligned}
			(\mathbf{R}_j)
			\left\{	\begin{array}{ll}
				(\mathcal{C}_{1j}) \hphantom{0} T_{{i}} \leq t_{{i}}^{{r}},  \forall i \in \Phi_j^{t+s},	\\
				(\mathcal{C}_{2j}) \hphantom{0} \sum_{i \in \Phi_j^{t}}r_{ij}^{f} \leq R_{j}^{f}, \\
				(\mathcal{C}_{3j}) \hphantom{0} \sum_{i \in \Phi_j^{t+s}}r_{ij}^{u} \leq R_{j}^{u}, \\
				(\mathcal{C}_{4j}) \hphantom{0} \sum_{i \in \Phi_j^{t+s}}r_{ij}^{d} \leq R_{j}^{d},\\
				(\mathcal{C}_{5j}) \hphantom{0} \sum_{i \in \Phi_j^{t+s}}b_{ij} \leq \mathcal{B}_{j},\\
				r_{ij}^{f}, r_{ij}^{u}, r_{ij}^{d}, b_{ij} \geq 0,\forall i \in \Phi_j^{t+s}, \\
				r_{ij}^{f} = 0,\forall i \in \Phi_j^{s}, ~
				b_{ij} = 0,\forall i \in \Phi_j^{t},
			\end{array}	\right.
		\end{aligned}
		\label{eq:resource_and_delay_constraints:TMC}
	\end{equation}
	where $(\mathcal{C}_{1j})$ captures the delay requirement of tasks, $(\mathcal{C}_{2j})$, $(\mathcal{C}_{3j})$, $(\mathcal{C}_{4j})$, and $(\mathcal{C}_{5j})$ are the computational, uplink, downlink, and backhaul constraints of EN~$j$.
	
	
	At iteration $(k)$, if $(M+1)$ subproblems $(\mathbf{SP}_1)$ are feasible at all $(M+1)$ ENs, then $\mathbf{x}^{(k)}$, \mbox{$\mathbf{r}=(\mathbf{r}_{1}, \dots, \mathbf{r}_{(M+1)})$}, and \mbox{$\mathbf{b}=(\mathbf{b}_{1}, \dots, \mathbf{b}_{(M+1)})$} are the optimal solutions of $(\mathbf{P}_0)$. Otherwise, for each infeasible $(\mathbf{SP}_1)$ at EN~$j$, a new Benders cut $c_j^{(k)}$, namely \textit{Subproblem Benders Cut}, will be added to the cutting-plane set of $(\mathbf{MP}_0)$ for the next iteration, i.e., $cuts^{(k+1)}=cuts^{(k)} \cup c_j^{(j)}$. These Benders cuts are designed in Section~\ref{sub:sub_cut:TMC}. 
	To further improve the efficiency of DBBD, we design the feasibility/infeasibility detection mechanism below. This mechanism is later used to generate Benders cuts to early remove infeasible solutions.
	
	\subsubsection{Feasibility and Infeasibility Detection}
	\label{sec:fast_feasible_infeasible:TMC}
	
	Replacing Eqs.~(\ref{eq:edge_delay:TMC}),~(\ref{eq:indirect_cloud_delay:TMC}),~and~(\ref{eq:direct_cloud_delay:TMC}) into $(\mathcal{C}_{1j})~T_{i} \leq t_i^r$ in $(\mathbf{R}_j)$ of $(\mathbf{SP}_1)$, this delay constraint can be transformed as
	\begin{equation}
		\begin{cases}			
			\left(\frac{L_{i}^{u}}{r_{ij}^{u}}+\frac{L_{i}^{d}}{r_{ij}^{d}}+\frac{L_{i}^{u}w_{i}}{r_{ij}^{f}}\right) \leq t_i^r - \zeta, \forall i \in \Phi_j^t , \\
			
			\left(\frac{L_{i}^{u}}{r_{ij}^{u}}+\frac{L_{i}^{d}}{r_{ij}^{d}}\right) + \left(\frac{L_{i}^{u}+L_{i}^{d}}{b_{ij}}\right) \leq t_i^r - 
			\frac{L_{i}^{u} w_{i}}{\mathcal{C}_{q}} - \zeta, \forall i \in \Phi_j^s.
		\end{cases}
		\label{eq:delay_contraint2:TMC}
	\end{equation}	
	Remarkably, $\left(t_i^r - \zeta\right)$ and $(t_i^r - 
	\frac{L_{i}^{u} w_{i}}{\mathcal{C}_{q}} - \zeta)$ are constant components.
	If $\exists i \in \Phi_j^{t+s}, (t_i^{r} - \zeta) \leq 0$ or $(t_i^r - \frac{L_{i}^{u} w_{i}}{\mathcal{C}_{q}} - \zeta) \leq 0$, then offloading task~$I_i$ to either EN~$j$ or the CS does not meet the delay requirement, i.e., $T_{i} \leq t_i^r$, leading to the infeasibility of $(\mathbf{SP}_1)$. In this case, a new cutting-plane is directly generated to prevent offloading task $I_i$.
	Otherwise, if $(t_i^{r} - \zeta) > 0, \forall i \in \Phi_j^{t}$ and $(t_i^r - \frac{L_{i}^{u} w_{i}}{\mathcal{C}_{q}} - \zeta) > 0, \forall i \in \Phi_j^{s}$, then the relative size, i.e., $(L_{i}^{u'}, L_{i}^{d'}, w_{i}^{'}, L_{i}^{c'})$, of task $I_i$ is defined as
	\begin{equation}
		\begin{cases}
			\left(\frac{L_{i}^{u}}{t_i^r - \zeta}, \frac{L_{i}^{d}}{t_i^r - \zeta}, w_i, 0 \right), & \forall i \in \Phi_j^t \\
			\left(\frac{L_{i}^{u}}{t_i^r - \frac{L_{i}^{u} w_{i}}{\mathcal{C}_{q}} - \zeta}, \frac{L_{i}^{d}}{t_i^r - \frac{L_{i}^{u} w_{i}}{\mathcal{C}_{q}} - \zeta}, 0, \frac{L_{i}^{u}+L_{i}^{d}}{t_i^r - \frac{L_{i}^{u} w_{i}}{\mathcal{C}_{q}} - \zeta} \right), & \forall i \in \Phi_j^s.
		\end{cases}
		\label{eq:define_task2:TMC}
	\end{equation}
	
	For task $I_i$, let $\beta_i = (
	\frac{L_{i}^{u'}}{r_{ij}^{u}}+
	\frac{L_{i}^{d'}}{r_{ij}^{d}}+
	\frac{L_{i}^{u'}w_{i}^{'}}{r_{ij}^{f}}+
	\frac{L_{i}^{c'}}{b_{ij}}
	)$. 
	Then, the delay constraint in Eq.~(\ref{eq:delay_contraint2:TMC}) becomes
	\begin{equation}
		\begin{aligned}
			\beta_i =
			\left(
			\frac{L_{i}^{u'}}{r_{ij}^{u}}+
			\frac{L_{i}^{d'}}{r_{ij}^{d}}+
			\frac{L_{i}^{u'}w_{i}^{'}}{r_{ij}^{f}}+
			\frac{L_{i}^{c'}}{b_{ij}}
			\right) \leq 1, \forall i \in \Phi_j^{t+s}.
		\end{aligned}
		\label{eq:delay_constraints_new2:TMC}
	\end{equation}
	
	Based on the relative size concepts, Theorems~\ref{theo:feasible:TMC}~and~\ref{theo:infeasible:TMC} below, respectively, can detect the feasibility and the infeasibility of $(\mathbf{SP}_1)$.
	
	\begin{theorem}
		\label{theo:feasible:TMC}
		Let  
		$\beta_{bal}^u = \frac{\sum_{i \in \Phi_j^{t+s}}
			L_i^{u'}}{R_j^u}$, $\beta_{bal}^d = \frac{\sum_{i \in \Phi_j^{t+s}}
			L_i^{d'}}{R_j^d}$, $\beta_{bal}^f = \frac{\sum_{i \in \Phi_j^{t+s}}
			L_i^{u'}w_i^{'}}{R_j^f}$, and $\beta_{bal}^b = \frac{\sum_{i \in \Phi_j^{t+s}}
			L_i^{c'}}{\mathcal{B}_j}$. If $\beta_{bal} = \beta_{bal}^u + \beta_{bal}^d + \beta_{bal}^f + \beta_{bal}^b \leq 1$, then $(\mathbf{SP}_1)$ is feasible and  $\mathbf{r}_{ij}=(r_{ij}^{u},r_{ij}^{d},r_{ij}^{f}, b_{ij})$ $ =  (\frac{L_i^{u'}}{\beta_{bal}^u},  \frac{L_i^{d'}}{\beta_{bal}^d}, \frac{L_i^{u'}w_i^{'}}{\beta_{bal}^f}, \frac{L_i^{c'}}{\beta_{bal}^b}),$ $\forall i \in \Phi_j^{t+s}$, is a resource allocation solution.

	\end{theorem}
	
	{\emph{Proof:}} The proof is shown in Appendix~\ref{sec:theo_feasible:TMC}.
	
	
	\begin{theorem}
		\label{theo:infeasible:TMC}
		If $\frac{\sum_{i \in \Phi_j^{t+s}}
			L_i^{u'}}{R_j^u} > 1$ or  $\frac{\sum_{i \in \Phi_j^{t+s}}
			L_i^{d'}}{R_j^d} > 1$ or  $\frac{\sum_{i \in \Phi_j^{t+s}}
			L_i^{u'}w_i^{'}}{R_j^f} > 1$ or $\frac{\sum_{i \in \Phi_j^{t+s}}
			L_i^{c'}}{\mathcal{B}_j} > 1$, then $(\mathbf{SP}_1)$ is infeasible.
	\end{theorem}
	
	{\emph{Proof:}} The proof is shown in Appendix~\ref{sec:theo_infeasible:TMC}.
	
	Theorem~\ref{theo:feasible:TMC} can detect the closed-form resource allocation solutions of $(\mathbf{SP}_1)$ at EN~$j$ without requiring an optimizer. As a result, the computation time is reduced. 
	Similarly, Theorem~\ref{theo:infeasible:TMC} can quickly detect the infeasibility of $(\mathbf{SP}_1)$ at EN~$j$. However, we will exploit Theorem~\ref{theo:infeasible:TMC} in a more effective way by developing Benders cuts, namely \textit{Resource Benders Cut}, for $(\mathbf{MP}_0)$ to prevent the generation of $(\mathbf{SP}_1)$ that volatile Theorem~\ref{theo:infeasible:TMC}. Those Benders cuts, presented in Section~\ref{sub:resource_cut:TMC}, will be added to the set $cuts$ of $(\mathbf{MP}_0)$ at the initial step of the DBBD. Consequently, a large number of useless offloading solutions are not generated, thus remarkably reducing the solving time of the DBBD.
	
	\subsubsection{Optimizing the Delay Satisfaction Rate}
	\label{sec:delay_satisfaction_rate:TMC}
	
	At EN~$j$, the fairness in terms of delay is considered. Particularly, we solve $(\mathbf{SP}_1)$ with the new form of delay constraint in Eq.~(\ref{eq:delay_constraints_new2:TMC}) so that all tasks gain the same delay satisfaction rate $\gamma_{j}$. The solution of $(\mathbf{SP}_1)$, which minimizes the delay satisfaction rate $\gamma_{j}$, is equivalent to those of its variant problem $(\mathbf{SP}_2)$ with an additional slack variable $\gamma_{j}$. 
	Here,  $(\mathbf{SP}_2)$ is always feasible.
	\begin{equation}
		(\mathbf{SP}_2) \hphantom{10}	 
		\underset{\mathbf{r}_j, \mathbf{b}_j, \gamma_{j}}{\min}\{\gamma_{j}\} , ~\text{s.t.}~(\mathbf{\widetilde{R}}_j),
		\label{eq:sub_constrant_goal_relax_var:TMC}
	\end{equation}
	\begin{equation}
		\begin{aligned}
			(\mathbf{\widetilde{R}}_j)
			\left\{	\begin{array}{ll}
				(\mathcal{C}_{1j}) \hphantom{1} \beta_i \leq \gamma_{j}, \forall i\in \Phi_j^{t+s},	\\
				(\mathcal{C}_{2j}) \hphantom{1} \sum_{i \in \Phi_j^{t}}r_{ij}^{f} \leq R_{j}^{f}, \\
				(\mathcal{C}_{3j}) \hphantom{1} \sum_{i \in \Phi_j^{t+s}}r_{ij}^{u} \leq R_{j}^{u}, \\
				(\mathcal{C}_{4j}) \hphantom{1} \sum_{i \in \Phi_j^{t+s}}r_{ij}^{d} \leq R_{j}^{d},\\
				(\mathcal{C}_{9j}) \hphantom{1} \sum_{i \in \Phi_j^{t+s}}b_{ij} \leq \mathcal{B}_{j},\\
				r_{ij}^{f}, r_{ij}^{u}, r_{ij}^{d} \geq 0,\forall i \in \Phi_j^{t+s}, \\
				r_{ij}^{f} \!=\! 0,\forall i \in \Phi_j^{s}, ~
				b_{ij}\!=\! 0,\forall i \in \Phi_j^{t}, 
				0 < \gamma_{j}.
			\end{array}	\right.
		\end{aligned}
		\label{eq:resource_and_delay_con:TMC}
	\end{equation}
	
	\begin{theorem}
		\label{theo:convexity_sub:TMC}
		The optimization problem $(\mathbf{SP}_2)$ is convex.
	\end{theorem}
	
	{\emph{Proof:}} The proof is shown in Appendix~\ref{sec:theo_convexity_sub:TMC}.
	
	Due to the convexity of $(\mathbf{SP}_2)$, we can use an optimizer to find its optimal resource allocation solution. If $(\mathbf{SP}_2)$ is feasible with the resulting delay satisfaction rate $\gamma_{j} \leq 1$, then $(\mathbf{SP}_1)$ is feasible with the resource allocation solution of $(\mathbf{SP}_2)$. Otherwise, we conclude the infeasibility of $(\mathbf{SP}_1)$.
	
	
	
	\subsection{Benders Cut Generation}
	\label{sub:Cutting_Plane_Generation:TMC}
	
	
	This section develops three types of Benders cuts, namely \emph{Subproblem Benders Cut}, \emph{Resource  Benders Cut}, and \emph{Prefixed-Decision Benders Cut}, that will be added to the constraints of $(\mathbf{MP}_0)$.
	Though the DBBD algorithm can find the optimal solution only by using the subproblem Benders cuts as presented below, the two other types of Benders cuts can help to reduce the search space, thereby significantly reducing the computation time of the algorithm. 
	This approach is more advanced than that proposed in~\cite{yu2018green}, in which only one Benders cut is updated at each iteration. 
	
	\subsubsection{Subproblem Benders Cut}\label{sub:sub_cut:TMC}
	At iteration $(k)$, the problem $(\mathbf{SP}_1)$ with assigned tasks $\Phi_j^{t+s}$ at EN~$j$ is determined by $\mathbf{x}_j^{(k)} = (\mathbf{x}_j^{f(k)},\mathbf{x}_j^{c(k)})$. If  $(\mathbf{SP}_1)$ is infeasible, then a new Benders cut $c_j^{(k)}$ will be added to the $cuts$ set of  $(\mathbf{MP}_0)$ to prevent offloading $\Phi_j^{t+s}$ in the next iterations.
	\begin{equation}
		c_j^{(k)}=\{\mathbf{x}_{j}^{f(k)\top} \mathbf{x}_{j}^{f} + \mathbf{x}_{j}^{c(k)\top} \mathbf{x}_{j}^{c} \leq t+s-1\}.
	\end{equation}

	\subsubsection{Resource Benders Cut}\label{sub:resource_cut:TMC}
	
	
{\color{black}	To guarantee a feasible problem $(\mathbf{SP}_1)$ at EN~$j$,  the set $\Phi_j^{t+s} \subseteq \Phi$, determined by  $(\mathbf{x}_j^f,\mathbf{x}_j^c)$, must not violate any resource constraints at EN~$j$ as stated in Theorem~\ref{theo:infeasible:TMC}. 
	Let \mbox{$\mathbf{c}_j^{u(edge)} \!=\! (L_1^{u'}, \dots, L_N^{u'} )/R_j^u$}, $\mathbf{c}_j^{d(edge)} = (L_1^{d'}, \dots, L_N^{d'})/R_j^d$ and $\mathbf{c}_j^{f(edge)} \!=\! (L_1^{u'}w_i^{'}, \dots, $ $L_N^{u'}w_i^{'})/R_j^f$. Here, $(L_i^{u'}, L_i^{d'}, w_i^{'})$ is defined as in Eq.~(\ref{eq:define_task2:TMC}) for $i \in \Phi_j^t$. 
	Let $\mathbf{c}_j^{u(cloud)} = (L_1^{u'}, \dots, L_N^{u'}) $ $/R_j^u$, $\mathbf{c}_j^{d(cloud)} = (L_1^{d'}, \dots, L_N^{d'})/R_j^d$, and $\mathbf{c}_j^{b(cloud)} = (L_1^{c'}, \dots, L_N^{c'})/\mathcal{B}_j$. Here, $(L_i^{u'}, L_i^{d'}, L_i^{c'})$ is defined as in Eq.~(\ref{eq:define_task2:TMC}) for $i \in \Phi_j^s$.
	
	To avoid the infeasible conditions in Theorem~\ref{theo:infeasible:TMC}, we add the cutting-planes below to the $cuts$ set of $(\mathbf{MP}_0)$. 
	$c_j^u=\{\mathbf{c}_j^{u(edge)\top} \mathbf{x}_j^f + \mathbf{c}_j^{u(cloud)\top}\mathbf{x}_j^{c} \leq 1\},$ 
	$c_j^d=\{\mathbf{c}_j^{d(edge)\top} \mathbf{x}_j^f + \mathbf{c}_j^{d(cloud)\top}\mathbf{x}_j^{c} \leq 1\},$ 
	$c_j^f=\{\mathbf{c}_j^{f(edge)\top} \mathbf{x}_j^f \leq 1\}, ~\text{and}~
	c_j^b=\{\mathbf{c}_j^{b(cloud)\top}\mathbf{x}_j^{c} \leq 1\}.$}
	

	\subsubsection{Prefixed-Decision Benders Cut}\label{sub:pre_cut:TMC}
	
	As 
	aforementioned in Section~\ref{sec:fast_feasible_infeasible:TMC}, 
	if $(t_i^{r} - \zeta) \leq 0$ and $(t_i^r - \frac{L_{i}^{u} w_{i}}{\mathcal{C}_{q}} - \zeta) \leq 0$, then task~$I_i$ cannot be offloaded to ENs and the CS, respectively. Thus, suitable cutting-planes can be generated and updated in the $cuts$ set of $(\mathbf{MP}_0)$. 
	From Table~\ref{tab:task_categories:TMC}, we also can add suitable Benders cuts according to their pre-decisions.
	
	\subsection{Solving the Master Problem}
	\label{sec:solving_master:TMC}
	
	To tackle $(\mathbf{MP}_0)$, we develop a low-complexity dynamic branch-and-bound algorithm, namely DBB, that can efficiently solve $(\mathbf{MP}_0)$ by exploiting the nature of binary decision variables as well as tasks' QoS requirements and ENs' available resources.
	Particularly, we first represent the computing offloading problem in the form of a decision tree. Each node on the tree is equivalent to a task, and the branches from that node are the possible processors (i.e., the UE, ENs, and the CS) toward the task.
	The computation/communication load of a branch is determined by the highest computation/communication load per processing unit amongst ENs/cloud. Then, the branches are arranged so that their load increases from left to right. Thus, a depth-first search will return the optimal solution satisfying the most load balancing amongst ENs and the cloud.
	In addition, an incremental search is implemented to speed up the optimal solution search processes by reusing the search results from the previous iterations (illustrated in Fig.~\ref{fig:DBBD_Model:TMC}). 
	In the following, we introduce the decision tree, the incremental depth-first search, the dynamic task selection, and the balancing processor selection.
	
	\subsubsection{Decision Tree of Tasks}
	\label{sec:branching_task:TMC}
	
	
	In the DBB algorithm, the decision tree has the following features.
	
	\begin{itemize}
		
		\item \label{Branching_Task:TMC} \textbf{Branching task}: 
		Assume that each computational task has $(2(M+1)+1)$ offloading choices, including $(M+1)$ edge nodes, $(M+1)$ cloud servers (via $(M+1)$ edge nodes), and one local device. In practice, the number of possible offloading choices can be less than $(2(M+1)+1)$ due to QoS requirements (e.g., delay, security). 
		Each node on the tree is equivalent to a task, and the branches from that node are the possible offloading choices for this task, forming a $(2(M+1)+1)$-tree with the depth of $(|\Phi|-1)$. The root node has a depth of $0$.
		
		
		\item \label{Simplifying_problem:TMC} \textbf{Simplifying problem}: 
		For the offloading variables $\mathbf{x}_i$ of task $i$, only one variable takes value $1$, whereas all others take value $0$ as in the offloading constraint $(\mathcal{C}_6)$ in Eq.~(\ref{eq:resources_and_offload_con_merge:TMC}). 
		Thus, if $x_{ij}^f = 0$ (also $x_{ij}^c = 0$), we can remove all expressions of the forms $x_{ij}^{f} A$ (also $x_{ij}^{c} B$), and these variables in~$(\mathbf{MP}_0)$.
		As a result, each node on the decision tree is also equivalent to an intermediate problem with fewer variables, namely $(\mathbf{IP})$, in which the ancestor nodes have tasks with the fixed offloading decisions.
		
		\item \label{Preserving_Convexity:TMC} \textbf{Preserving convexity}: Let $(\mathbf{\widetilde{MP}}_0)$ and $(\mathbf{\widetilde{IP}})$, respectively, be the relaxed problems of $(\mathbf{MP}_0)$ and $(\mathbf{IP})$. Due to the convexity of $(\mathbf{\widetilde{MP}}_0)$, the problem $(\mathbf{\widetilde{IP}})$, which is equivalent to $(\mathbf{\widetilde{MP}}_0)$ with some fixed offloading variables, is convex.
	\end{itemize}
	
	The low complexity of the DBB algorithm with the above characteristics compared with the conventional branch-and-bound approach is evaluated in Section~\ref{sec:Complexity_Analysis:TMC}.

	\subsubsection{Incremental Depth-First Search}
	\label{sec:incremental_search:TMC}
	
	
	If we consider all possible offloading policies as a search space, then the search space will be partitioned into subspaces by intermediate problems of the form $(\mathbf{IP})$ mentioned in Section~\ref{sec:branching_task:TMC}. Here, we will introduce the incremental depth-first search (namely IDFS) to find the optimal offloading policy of $(\mathbf{MP})$.	
	At iteration~$(k)$, at a node of $(\mathbf{IP})$ on the tree, the result of the relaxed problem $(\mathbf{\widetilde{IP}})$ will be evaluated to determine the potential of having an optimal solution in that subspace. Then, a suitable action, i.e., branching or pruning, will be carried out.  
	In the IDFS, the results from previous iterations will be evaluated so that a new search will be carried out on the sub-tree from that node only if it is an undiscovered potential subspace. Consequently, the solving time can be significantly reduced. There are two cases with the relaxed problem $(\mathbf{\widetilde{IP}})$ as follows.
	
	
	\textbf{Case 1:} If $(\mathbf{\widetilde{IP}})$ was feasible at previous iterations and the best result of both $(\mathbf{\widetilde{IP}})$ and $(\mathbf{IP})$ (if it exists) of previous iterations is not greater than that of the current optimal one, then we will prune the sub-tree starting from the node of $(\mathbf{IP})$ at the current iteration $(k)$, and the sub-tree will be stored for evaluation in future iterations.
	
	\textbf{Case 2:} If either $(\mathbf{\widetilde{IP}})$ was not solved or $(\mathbf{\widetilde{IP}})$ was feasible at the previous iterations and the best result of both $(\mathbf{\widetilde{IP}})$ and $(\mathbf{IP})$ (if it exists) from previous iterations is greater than that of the current optimal one, then $(\mathbf{\widetilde{IP}})$ will be solved to determine the potential of having an optimal solution in that subspace. There are three possibilities when solving $(\mathbf{\widetilde{IP}})$ as below.

	\begin{itemize}
		
		\item \label{ip_infeasible:TMC} If $(\mathbf{\widetilde{IP}})$  is infeasible, then we will prune the sub-tree starting from the node of $(\mathbf{IP})$ at the current iteration $(k)$ and future iterations.
		
		\item \label{ip_feasible_nobetter:TMC} If $(\mathbf{\widetilde{IP}})$  is feasible and the result is not better than that of the current optimal one, then we will prune the sub-tree starting from the node of $(\mathbf{IP})$ at the current iteration $(k)$, and the sub-tree will be stored for evaluation in future iterations.
		
		\item \label{ip_feasible_better:TMC} If $(\mathbf{\widetilde{IP}})$ is feasible and the result is better than that of the current optimal one, and if this node was not branched in previous iterations, we will choose a task to branch at this node as shown in Sections~\ref{sec:dynamic_task_selection:TMC}~and~\ref{sec:balancing_processor_selection:TMC}. The relaxed solution of $(\mathbf{\widetilde{IP}})$ will be updated as the current optimal if it is an integer solution.

		
	\end{itemize}

	To reuse results between iterations, the structure of the decision tree needs to inherit and expand from previous iterations. Thus, the results at a node are correlated and comparable between iterations. The tree is also designed flexibly so that the global optimal solution can be quickly found in each iteration. We thus develop the dynamic task and processor selection policies in the following sections, which are applied to the undiscovered portion of the search space in iterations.

	\subsubsection{Dynamic Task Selection}
	\label{sec:dynamic_task_selection:TMC}
	
	The DBB algorithm travels through the decision tree to find the optimal solution by constantly updating the current solution with better ones. All the branches with no better solutions will be pruned without traveling. Thus, the sooner a better solution (i.e., the solution is close to the optimal one) is found, the more sub-spaces of the tree are pruned without traveling. Hence, it significantly reduces the solving time.  
	
	
	The tasks in $\widetilde{\Phi}$ always need to be offloaded to satisfy their requirements, whereas a proportion of tasks in $\widehat{\Phi}$ may not be offloaded due to ENs' resource limitation. Thus, in the DBB, the tasks in $\widetilde{\Phi}$ will be  chosen before the ones in $\widehat{\Phi}$. 
	Besides, the tasks with higher energy benefits per required resource unit are likely offloaded to ENs/cloud in the optimal solution. Thus, in each group, i.e., $\widetilde{\Phi}$ and $\widehat{\Phi}$, , these tasks will be early chosen at nodes close to the root of the tree. 
	
	From Eq.~(\ref{eq:task_en:TMC}), we can determine the maximum benefits of task~$I_i$ among all possible offloading decision solutions as $\Delta_{i}^{max}= \underset{\mathbf{x}_{i}}{\max} \{\Delta_{i}\}$.
	Then, from Eqs.~(\ref{eq:e_base:TMC})~and~(\ref{eq:task_en:TMC}), we have
	\begin{equation}
		\Delta_{i}^{max}=\begin{cases}
			E_{i}^{l} - \underset{j \leq (M+1) }{\min}\{E_{ij}^f\}, & I_i \in \widehat{\Phi}, \\
			\underset{j \leq (M+1) }{\max}\{E_{ij}^f\} - \underset{j \leq (M+1) }{\min}\{E_{ij}^f\}, & I_i \in \widetilde{\Phi} .
		\end{cases}\label{eq:delta_max:TMC}
	\end{equation}
	
		
		For simplicity, we assume that the delays of tasks mostly belong to the computation.  
		Let $r_i^{min} = \frac{L_i^u w_i}{t_i^r - \zeta}$ be a lower bound of computation resource required by task $I_i$. 
		We then define the rate of benefits as $rate_i = \frac{\Delta_{i}^{max}}{r_i^{min}}$. 

For each UE~$n$ with the task set $\Phi_{n}$ ($1 \leq n \leq N$), let $\Phi^{+}_{n}$ be the set of tasks chosen at the ancestors of the current node. Let $\Phi^{*}$ be the set of $N$ tasks with the highest rate $rate_i$ from $N$ UEs. We have 
\begin{equation}
\Phi^* = \{I_n~|~I_n = \underset{I_i \in \Phi_n / \Phi_{n}^{+} }{\argmax} rate_i, \forall n \leq N\}.
\label{eq:select_n_task:TMC}
\end{equation}


Here, the computational task $I_{i^*}$ will be selected if it can help to increase the utility function most. Thus, the selection of task $I_{i^*}$ can be determined as:
\begin{equation}
\label{eq:select_task_1:TMC}
\begin{split}
I_{i^*} &\!=\! \underset{I_{i} \in \Phi^{*}}{\argmax} \left(\rho_i \ln \left(u_{n}^{+}+\Delta_{i}^{max} \right) - \rho_i \ln \left(u_{n}^{+} \right)\right)  \\
&\!=\! \underset{I_{i} \in \Phi^{*}}{\argmax}~ \rho_i \ln \left(1 + \frac{\Delta_{i}^{max}}{u_{n}^{+}} \right), 
\end{split}
\end{equation}
in which $u_{n}^{+}$ is the total utility of all tasks in $\Phi_{n}^{+}$. 



\subsubsection{Balancing Processor Selection}
\label{sec:balancing_processor_selection:TMC}

For a selected computational task $I_{i^*}$ as in Section~\ref{sec:dynamic_task_selection:TMC}, we choose a processor (i.e., the UE, an EN, or the CS), aiming to balance the joint communication and computation load amongst ENs and the CS.
Mathematically, this helps to create the most efficient subproblem Benders cuts at early iterations.
Consequently, the optimal offloading solution satisfying the feasible resource allocation at all ENs can be found with a few iterations, thereby reducing the overall solving time. Besides, the delay satisfaction rate as in Section~\ref{sec:delay_satisfaction_rate:TMC} will be balanced and optimized amongst the tasks. 



At the current node of the depth $l$ of the decision tree, assume that task $I_{i^{*}}$ is selected. 
We need to sort the possible processors (i.e., the UE, ENs, and the CS) towards task $I_{i^{*}}$ in the ascending order of estimated delay. 
Let $\Phi_j^{t^*}$ and $\Phi_j^{s^*}$, respectively, be the temporary sets of chosen tasks being processed at EN~$j$ and the CS (via EN~$j$). 
Let $\beta_{j}^{f}$ and $\beta_{j}^{c}$ be the upper bounds of the delay satisfaction rate of all tasks in $\Phi_j^{t^*+s^*}$ and $I_{i^{*}}$ when they are executed at EN~$j$ and the CS (via EN~$j$), respectively. 
We define these parameters as follows.
\begin{equation}
\label{eq:select_node:TMC}
\begin{split}
\beta_{j}^{f} = &\frac{\sum \limits_{i \in \Phi_j^{t^*+s^*}}
	L_i^{u'} + L_{i^{*}}^{u'}}{R_j^u} + 
\frac{\sum \limits_{i \in \Phi_j^{t^*+s^*}}
	L_i^{d'} + 
	L_{i^{*}}^{d'}}{R_j^d} \\
&+ \frac{\sum \limits_{i \in \Phi_j^{t^*}}
	L_i^{u'}w_{i}^{'} + L_{i^{*}}^{u'} w_{i^{*}}^{'}}{R_j^f}
, \forall j \in \mathbb{G}(q),			
\end{split}
\end{equation}
\begin{equation}
\begin{split}
\beta_{j}^{c} = & \frac{\sum \limits_{i \in \Phi_j^{t^*+s^*}}
	L_i^{u'} + L_{i^{*}}^{u'}}{R_j^u} + 
\frac{\sum \limits_{i \in \Phi_j^{t^*+s^*}}
	L_i^{d'} + L_{i^{*}}^{d'}}{R_j^d} \\
&+ \frac{\sum \limits_{i \in \Phi_j^{t^*}}
	L_i^{c'} + L_{i^{*}}^{c'} }{\mathcal{B}_j}
,
\end{split}
\label{eq:select_cloud:TMC}
\end{equation}
where $\mathbb{G}(q)$ is the set of all ENs that support the application type $q$.	
Each task $I_{i^{*}}$ will be then offloaded to the processors in the preference order of $\left(F_{1}, \dots, C_{1}, \dots, L\right)$. $\beta_{j}^{f} \leq \beta_{j+1}^{f}$ and $\beta_{j}^{c} \leq \beta_{j+1}^{c}$, in which $F_{j}$, $C_{j}$ and $L$, respectively, determine processors as EN~$j$, the CS (via EN~$j$), and the UE towards task $I_{i^{*}}$.

%

In Algorithm~\ref{DBB_Algorithm_code:TMC}, the DBB is structured as a decision tree as in Section~\ref{sec:branching_task:TMC}. At every node on the tree, the most suitable task is selected for branching as in Section~\ref{sec:dynamic_task_selection:TMC}, then the branches from this node are developed according to the order of processors as in Section~\ref{sec:balancing_processor_selection:TMC}. To find the optimal offloading solution, the DBB algorithm travels through the tree between nodes via edges determined by the branches, using the incremental depth-first search as in Section~\ref{sec:incremental_search:TMC}.  
The proposed DBBD algorithm, which uses the DBB to solve the master problem, is introduced as follows.

{\renewcommand{\baselinestretch}{1}
\begin{algorithm} \small
	\DontPrintSemicolon
	\SetKwInput{Left}{left}\SetKwData{This}{this}\SetKwData{Up}{up}
	\SetKwInOut{Input}{Input}\SetKwInOut{Output}{Output}
	
	\Input{Set $\Phi$ of tasks $I_{i}\left(L_{i}^{u},L_{i}^{d},w_{i},t_{i}^{r}, s_{i}^{r}, q, n \right)$; Set of $M$ ENs $\{(R_{j}^{u},R_{j}^{d},R_{j}^{f}, s_{j}^{f}, \varPsi_j)\}$; Set $\mathcal{N}$ of UEs \\  Security levels $\mathbb{S}$; Application types $\mathbb{Q}$; Cloud server $(\mathcal{B}, \mathcal{C})$; Decision tree at $k$-th iteration $tree^{(k)}$} 
	\Output{Optimal $(\mathbf{x}^{(k)},maxU)$ of $(\mathbf{MP}_0)$; 
		Decision tree for next iteration $tree^{(k+1)}$}
	\BlankLine
	
	\Begin{
		$(\mathbf{x}^{(k)},maxU) \gets ( \emptyset,-\infty)$; 
		$tree^{(k+1)}.empty()$ \Comment{Empty solution and empty tree for next iteration}\;
		\lIf{$k = 1$}{
			$tree^{(k)}.push((\mathbf{MP}_0))$
		} 
		\While{$tree^{(k)}.isNotEmpty()$}{
			$p \gets tree^{(k)}.pop()$\Comment{Get $(\mathbf{{IP}})$ from top of stack}\;
			\uIf{$p$ \textnormal{was solved \textbf{and} its result is not better than} $maxU$}{		
				
				$tree^{(k+1)}.push(\textnormal{sub-tree from}~ p)$ \Comment{Store all $(\mathbf{{IP}})$ on sub-tree from $p$ for next iteration}\;
				
				\underline{Prune} sub-tree from $p$ in $tree^{(k)}$\;
				
				\textbf{continue} \Comment{Skip $p$ at current iteration}\;
			}
			
			$(\widetilde{\mathbf{x}},subU) \gets$ \underline{Solve} $(\mathbf{\widetilde{IP}})$ of $p$ then return its relaxed optimal solution and value\;
			
			\uIf{$\widetilde{\mathbf{x}}$ \textnormal{is not found}}{ 
				\underline{Prune} sub-tree from $p$ in $tree^{(k)}$\;
			}\uElseIf{$subU \leq maxU$}{
				$tree^{(k+1)}.push(\textnormal{sub-tree from}~ p)$; 
				\underline{Prune} sub-tree from $p$ in $tree^{(k)}$\;
			}			
			\uElseIf{$\widetilde{\mathbf{x}}$ \textnormal{are integer}}{
				$(\mathbf{x}^{(k)},maxU) \gets ( \widetilde{\mathbf{x}},subU)$
				\Comment{Update solution and optimal result}\;
				$tree^{(k+1)}.push(\textnormal{sub-tree from}~ p)$; 
				\underline{Prune} sub-tree from $p$ in $tree^{(k)}$\;					
			}\uElseIf{$p$ \textnormal{was not branched}}{
				Find task $I_{i^*}$ and processors $\left(F_{1^{*}}, \dots, C_{1^{*}}, \dots, L\right)$ for $I_{i^*}$ as in Sections~\ref{sec:dynamic_task_selection:TMC}~and~\ref{sec:balancing_processor_selection:TMC}\;
				$list \gets$ \underline{Branch} $p$ to create sub-tree of $p$ by trying decisions of  $I_{i^*}$ in order $\left(F_{1^{*}}, \dots, C_{1^{*}}, \dots, L\right)$\;
				\For{{\bf{each}} $chil$ {\bf in} $list$}{
					\underline{Simplify} $chil$ as in Section~\ref{Simplifying_problem:TMC}; 
					$tree^{(k)}.push(chil)$ \Comment{Put problem into stack}\; 
				}
				$tree^{(k+1)}.push(\textnormal{sub-tree from}~ p)$ \Comment{For next iteration}\;
			}			
		}
		
		\textbf{Return} $(\mathbf{x}^{(k)},maxU)$ and $tree^{(k+1)}$\;
	}
	\caption{DBB Algorithm\label{DBB_Algorithm_code:TMC}}\vspace{-.0in}
\end{algorithm}
}





\subsection{DBBD Algorithm}

The DBBD algorithm, presented in \textbf{Algorithm}~\ref{DBBD_Algorithm_code:TMC}, finds the optimal solution by iteratively solving $(\mathbf{MP}_0)$ and $(\mathbf{SP}_1)$. 
Initially, it initializes the iterator $k\!=\!1$ and sets $cuts^{(k)}$ in $(\mathbf{MP}_0)$ with $4(M\!+\!1)$ resource Benders cuts as in  Section~\ref{sub:resource_cut:TMC}. Other prefixed-decision Benders cuts are also added to $cuts^{(k)}$ as in Section-\ref{sub:pre_cut:TMC}.
At iteration $(k)$, the DBB algorithm finds $\mathbf{x}^{(k)} \in X_0$ of $(\mathbf{MP}_0)$ satisfying $cuts^{(k)}$. With $\mathbf{x}^{(k)}$, $(M+\!\!1)$ problems of the form $(\mathbf{SP}_1)$ with assigned tasks $\Phi_j^{s+t} \!\subseteq\! \Phi$ at $(M \!+\!1)$ ENs are defined. 
Then, using a convex solver, every EN~$j$ independently solves the variant of $(\mathbf{SP}_1)$, i.e., $(\mathbf{SP}_2)$, to find a resource allocation solution toward $\Phi_j^{t+s}$. 
Before that, Theorem~\ref{theo:feasible:TMC} can determine the feasibility of $(\mathbf{SP}_1)$. If $(\mathbf{SP}_1)$ has no solution, a new Benders cut $c_j^{(k)}$ as in Section~\ref{sub:sub_cut:TMC} will be updated into $cuts^{(k+1)}$ of $(\mathbf{MP}_0)$ for the later iterations. 
If $\mathbf{x}^{(k)}$ of $(\mathbf{MP}_0)$ does not exist, then DBBD can conclude the infeasibility of $(\mathbf{P}_0)$. With $\mathbf{x}^{(k)}$ of $(\mathbf{MP}_0)$, if $(M\!+\!1)$ problems of the form $(\mathbf{SP}_1)$ have solutions \mbox{$\left(\mathbf{r}, \mathbf{b}\right) \!=\! \left(\{\mathbf{r}_j\},\{\mathbf{b}_j\}\right)$}, then DBBD can conclude $(\mathbf{x}^{(k)}, \mathbf{r}, \mathbf{b})$ is the optimal solution of $(\mathbf{P}_0)$. 

In Algorithm~\ref{DBBD_Algorithm_code:TMC}, Theorem~\ref{theo:feasible:TMC} is used to check the feasibility of $(\mathbf{SP}_1)$ before calling the solver. Additionally, by using Theorem~\ref{theo:infeasible:TMC}, the resource cutting-planes are created at the initial stage of $(\mathbf{MP}_0)$. Thus, the subproblems violating Theorem~\ref{theo:infeasible:TMC} are prevented during the iterations. As a result, the computation time of the DBBD algorithm can be remarkably reduced.

{\renewcommand{\baselinestretch}{1}
	\begin{algorithm} \small
		\DontPrintSemicolon
		\SetKwInput{Left}{left}\SetKwData{This}{this}\SetKwData{Up}{up}
		\SetKwInOut{Input}{Input}\SetKwInOut{Output}{Output}
		\Input{Set $\Phi$ of tasks $I_{i}\left(L_{i}^{u},L_{i}^{d},w_{i},t_{i}^{r}, s_{i}^{r}, q, n \right)$; Set of $(M+1)$ ENs $\{(R_{j}^{u},R_{j}^{d},R_{j}^{f}, s_{j}^{f}, \varPsi_j)\}$\\ Set $\mathcal{N}$ of UEs, Security levels $\mathbb{S}$; Application types $\mathbb{Q}$, Cloud server $(\mathcal{B}, \mathcal{C})$}
		\Output{Optimal  $(\mathbf{x},\mathbf{r},\mathbf{b})$ of  $(\mathbf{P}_0)$}
		\BlankLine
		\Begin{
			$k \gets (k+1)$;  $cuts^{(k)} \gets \bigcup_{j=1}^{M+1} \{c_j^u, c_j^d, c_j^f, c_j^b\}$.\;
			
			\While{\textnormal{solution} $(\mathbf{x},\mathbf{r},\mathbf{b})$ \textnormal{has not been found}}{
				
				$\mathbf{x} \gets$ \textbf{DBB algorithm} \underline{solve} $(\mathbf{MP}_0)$ with $cuts^{(k)}$. \Comment{$\mathbf{x}$ stores $\mathbf{x}^{(k)}$ at iteration $k$}\;
				
				\uIf{$\mathbf{x}$\textnormal{ is found}}{
					Solution $\mathbf{x}$ defines $(M+1)$ problems $(\mathbf{SP}_1)$ with asigned tasks $\Phi_1^{t+s}, \dots \Phi_{M+1}^{t+s}$.
				}\lElse{
					\textbf{Return} Problem $(\mathbf{P}_0)$ is infeasible.
				}
				
				\For{ $(j = 1;\ j \leq M+1;\ j = j + 1)$}{
					$(\mathbf{r}_j, \mathbf{b}_j) \gets $ Solver \underline{solves} $(\mathbf{SP}_1)$ at EN~$j$ with assigned tasks $\Phi_j^{t+s}$.\;
					
					\uIf{$(\mathbf{r}_j, \mathbf{b}_j)$ \textnormal{is not found}}{
						
						Update new cut $c_j^{(k)}$ into $cuts^{(k+1)}$. 
					}		
				}
				
				\uIf{$\left(\mathbf{r}, \mathbf{b}\right) = \left(\{\mathbf{r}_j\},\{\mathbf{b}_j\}\right)$\textnormal{ is found}}{
					Optimal $(\mathbf{x},\mathbf{r},\mathbf{b})$  has been found.\;
				}
				$k \gets (k+1)$ \Comment{For next  iteration}\;
			}
			\textbf{Return} $(\mathbf{x},\mathbf{r},\mathbf{b})$\;
		}
		\caption{DBBD Algorithm
			\label{DBBD_Algorithm_code:TMC}}
	\end{algorithm}
}
\vspace{-.2in}



\subsection{Complexity Analysis}
\label{sec:Complexity_Analysis:TMC}


In this section, we analyze the complexity of the DBBD w.r.t. the number of tasks and ENs.

\subsubsection{Size of Original Problem}
With $(M+1)$ ENs including the cloud server $V$, the original problem $(\mathbf{P}_0)$ has $4(M+1)$ resource constraints for $(\mathcal{C}_2), (\mathcal{C}_3), (\mathcal{C}_4),~\text{and}~ (\mathcal{C}_5)$ described in Eq.~(\ref{eq:resources_and_offload_con_merge:TMC}). In addition, to formulate each task $I_i$ in $(\mathbf{P}_0)$, we need to consider $(2(M+1)+1)$ binary and $4(M+1)$ real variables, together with three constraints for the delay, offloading, and security as shown in~$(\mathcal{C}_1)$, $(\mathcal{C}_6)$~and~$(\mathcal{C}_7)$. The constraints $(\mathcal{C}_8)$~and~$(\mathcal{C}_9)$ in Eq.~(\ref{eq:resources_and_offload_con_merge:TMC}) fix the offloading variables, and thus they are not counted here. 
Therefore, with $|\Phi|$ tasks and $(M+1)$ edge nodes, the original problem $(\mathbf{P}_0)$ has respectively $|\Phi|(2(M+1)+1)$ integer and $4|\Phi|(M+1)$ real variables, and $(3|\Phi|+4(M+1))$ constraints including $|\Phi|$ for the offloading decisions, $|\Phi|$ for the security requirements, $|\Phi|$ for the delay requirements of the tasks and $4(M+1)$ for the resource requirements of ENs and cloud as described in Eq.~(\ref{eq:resources_and_offload_con_merge:TMC}).
Thus, the relaxed problem $(\mathbf{\widetilde{P}}_0)$ of $(\mathbf{P}_0)$ has totally $|\Phi|(6(M+1)+1)$ real variables and $(3|\Phi|+4(M+1))$ constraints.
\subsubsection{Size of Problems in DBBD}
For the DBBD algorithm, the master problem $(\mathbf{MP}_0)$ and $(M+1)$ subproblems of the form $(\mathbf{SP}_2)$ are iteratively solved. At iteration $k$, $(\mathbf{MP}_0)$ is an integer problem with $|\Phi|(2(M+1)+1)$ binary offloading variables and at most $(2|\Phi|+4(M+1) + k(M+1)$ constraints including $2|\Phi|$ for the offloading decision and security requirements as  $(\mathcal{C}_5)$~and~$(\mathcal{C}_7)$ described in Eq.~(\ref{eq:resources_and_offload_con_merge:TMC}), $4(M+1)$ for resource Benders cuts as in Section~\ref{sub:resource_cut:TMC}, and at most $k(M+1)$ for the subproblem Benders cuts from solving $(M+1)$ subproblems $k$ times as in Section~\ref{sub:sub_cut:TMC}. Furthermore, each subproblem $(\mathbf{SP}_2)$ is assigned an average of $|\Phi|/(M+1)$ tasks. Thus, it has approximate $4|\Phi|/(M+1)$ variables and $(|\Phi|/(M+1)+4)$ constraints including $|\Phi|/(M+1)$ for the delay of $|\Phi|/(M+1)$ tasks as $(\mathcal{C}_{1j})$ in Eq.~(\ref{eq:resource_and_delay_constraints:TMC}) and $4$ constraints for the resources requirements at the edge node as $(\mathcal{C}_{2j})$, $(\mathcal{C}_{3j})$, $(\mathcal{C}_{4j})$, and $(\mathcal{C}_{9j})$ described in Eq.~(\ref{eq:resource_and_delay_constraints:TMC}). In the worst case, $(\mathbf{SP}_2)$ is assigned all $|\Phi|$ tasks, and thus it has at most $4|\Phi|$ resources allocation variables and $(|\Phi|+4)$ constraints. However, if this big subproblem violates the resources constraints at the edge node according to Theorem~\ref{theo:infeasible:TMC}, it will not be created due to the generation of resource Benders cuts as in Section~\ref{sub:resource_cut:TMC}. 
\subsubsection{Complexity of DBBD}
With $|\Phi|$ tasks and $M$ ENs, there are $M^{|\Phi|+1}$ possible subproblems  $(\mathbf{SP}_1)$ with the task numbers increasing from $0$ to $|\Phi|$ and $M^{|\Phi|}$ master problems $(\mathbf{MP}_0)$. In the worst case, the DBBD has complexity in the order of $O(M^{|\Phi|})$.
However, with the support of Benders cut generations, most of the useless subproblems are excluded. 
Thus, in practice, the solving time is far less than that of the worst case. This is also because $(\mathbf{MP}_0)$ and $(\mathbf{SP}_1)$ have linear sizes over the number of tasks. {\color{black}The extensive simulations in the following section confirms the above analysis.}

\section{Performance Evaluation} 
\label{sec:performanceevaluation:TMC}

\subsection{Fairness Metrics} 

In edge networks, a large number of UEs/tasks often interact and share the same communication/ computation resources of ENs/cloud. Thus, we will study how the numbers of UEs/tasks and the available resources at ENs affect the fairness, energy benefits, and total consumed energy of all UEs. 
We will use Jain's index and min-max ratio to capture the fairness~\cite{jain1984quantitative}. These indexes are defined as $\textbf{Jain's index} = \frac{\left(\sum_{n=1}^{N}  u_n\right)^2}{N \sum_{n=1}^{N} u_n^2}$ and $\textbf{Min-Max ratio} = \frac{\underset{n \leq N }{\min}\{u_n\}}{\underset{n \leq N }{\max}\{u_n\}}$ where $u_n$ is the utility function of devices~$n$ as in Eq.~(\ref{eq:utility_fuc_:TMC}). 



\subsection{Experiment Setup}


{\color{black}
In this paper, to carry out the simulations, we first prepare the experiment data containing information about fog, cloud, UEs, and computational tasks with properties as described in Section~\ref{sec:sysmodel_sub:TMC}.
We create scenarios to evaluate how different aspects (i.e., resource limitation and problem sizes) affect the fairness and energy efficiency. 
For the optimization, we then implement the proposed method and benchmarks using the \textbf{MOSEK} Optimizer API~\cite{mosek2019documentation}. Finally, we run these algorithms with experiment data and analyze the results. 
}

We adopt the configuration of the Nokia N900 for all UEs as presented in \cite{miettinen2010energy}. 
The connections between UEs and ENs are either the 3G near connections or WLAN connections of the Nokia N900 as in Table~2~in~\cite{miettinen2010energy}. For the WLAN connections, we vary $e_{ij}^{u}$ and $e_{ij}^{d}$ in ranges $[0.5,1.5]$ times of standard values (i.e., $0.142$ J/Mb and $0.142$ J/Mb) described in Table~2~in~\cite{miettinen2010energy}. 
In this work, all UEs have the same weight $\rho_n = 1$, which shows that they are considered equally and fairly. 
Based on the computational task of the face recognition application as in \cite{chen2016efficient}, we then generate tasks $I_{i}\left(L_{i}^{u},L_{i}^{d},w_{i},t_{i}^{r}, s_{i}^{r}, q, n \right)$ $\left(I_i \in \Phi \right)$ as $L_{i}^{u} = 1$~MB,  $L_{i}^{d} =0.1$~MB, $w_{i} =5$~Giga~cycles/Mb, $t_{i}^{r} = 5$s, $s_{i}^{r} \in \mathcal{S}$, and $q \in \mathcal{Q}$. 
Three ENs have the uplink and downlink capacity surrounding $72$~Mbps, the highest WiFi theoretical  physical-layer data rate of 802.11n smartphones~\cite{liu2015small,saha2015power}. 
Besides, while each EN can randomly support only $3$ application types, the CS can support every application type in $\mathcal{Q}$. 
Other parameters are given in Table~\ref{tab:Experimental-parameters:TMC}. Different settings are provided in specific experiment scenarios. 

	{\renewcommand{\baselinestretch}{1}	
		\begin{table*}[htbp]
			\caption{Experimental parameters\label{tab:Experimental-parameters:TMC}} 
			\centering{}%
			\footnotesize
			\begin{tabular}{|l|c|l|c|}
				\cline{1-4}
				\textbf{Parameters} & \textbf{Value}
				&\textbf{Parameters} & \textbf{Value} \tabularnewline
				\cline{1-4} 
				
				Number of UEs $N$ & $2$ -- $12$ 
				& Application type $q \in \mathbb{Q}$ & $\mathbb{Q}=\{1,\ldots,5\}$ \tabularnewline
				\cline{1-4}
				
				Number of ENs $M$ & $3$ 
				& Processing rate of each EN $R_j^f$ & $1.5$ -- $15$ Giga cycles/s \tabularnewline
				\cline{1-4}
				
				Number of computation tasks $|\Phi|$ & $2$ -- $24$ 
				& Uplink rate of each EN $R_j^u$ & $11$ -- $110$ Mbps \tabularnewline
				\cline{1-4}
				
				CPU rate of UEs $f_i$ & $1$ Giga cycles/s 
				& Downlink rate of each EN $R_j^d$ & $11$ -- $110$ Mbps \tabularnewline
				\cline{1-4} 
				
				Security level of UEs $s_i$ & $1 (\text{High})$ 
				& Security level of each EN $s_j^f \in \mathbb{S}$ & $\mathbb{S}=\{1,\ldots,3\}$ \tabularnewline
				\cline{1-4}
				
				Energy model of UEs $(\alpha, \gamma)$ & $(10^{-11} \text{Watt/cycle}^{2}, 2)$ 
				& CPU rate of cloud $\mathcal{C} = \{\mathcal{C}_{1},\ldots,\mathcal{C}_{Q}\}$ & $\{10, \ldots, 10\}$ Giga cycles/s \tabularnewline
				\cline{1-4}
				
				Enery consumption rates of & $(0.071 - 0.213,$  &   Backhaul capacity $\mathcal{B} = \{\mathcal{B}_{1},\ldots,\mathcal{B}_{M}\}$ & $\{100, \ldots, 100\}$ Mbps  \tabularnewline
				\cline{3-4}
				
				WLAN connections $(e_{ij}^u,e_{ij}^d)$ & $0.071 - 0.213)$ J/Mb & Upper bound of backhaul rate & $b_{ij} \leq 5$ Mbps\tabularnewline
				\cline{1-4} 
				
				Enery consumption rates of & $(0.658,~0.278)$ J/Mb & Security level of cloud $s_{q}^c \in \mathbb{S}$ & $\mathbb{S}=\{1,\ldots,3\}$ \tabularnewline
				\cline{3-4}
				
				3G near connections $(e_{ij}^u,e_{ij}^d)$ &  & Multi-access delay $\zeta$ & $20$ms \tabularnewline
				\cline{1-4}
				
			\end{tabular}
		\end{table*}
	}
	
	
 { \color{black}
    Note that this work aims to develop an energy-based proportionally fair task offloading and resource allocation framework, i.e., DBBD, for a multi-layer cooperative edge computing network to serve all UEs while considering both their service requirements and individual energy/battery levels.  
    As such, one of the selected baselines is the total utility maximization framework, named FFBD, where the total energy consumption is minimized.
    }
 This benchmark is also called the social welfare maximization scheme (SWM) in~\cite{nguyen2018price,nguyen2019market}. The DBBD and FFBD are implemented using the \textbf{MOSEK} Optimizer API~\cite{mosek2019documentation}. 
	To highlight other performance (i.e., load balance and average delay) of the DBB algorithm that helps the DBBD in solving the MP, we implement two Benders decomposition variants of the FFBD, namely FFBD-I and FFBD-B. The FFBD-I uses the default linear \textbf{MOSEK} integer solver to solve the MP, whereas the FFBD-B solves the MP using a conventional branch-and-bound method without the load balancing implementation. In the FFBD-B, tasks are offloaded to EN~$j$ as much as possible before EN~$j+1$. 
	In the DBBD and FFBD, the minimum and maximum energy benefits of UEs are denoted by $\textnormal{DBBD}^{\textnormal{min}}$, $\textnormal{DBBD}^{\textnormal{max}}$ and $\textnormal{FFBD}^{\textnormal{min}}$,  $\textnormal{FFBD}^{\textnormal{max}}$, respectively. 
	We compare different schemes using the same data sets and capture the main trends in the figures. 
	
	\subsection{Numerical Results} 
	
	
	\subsubsection{Scenario~\ref{sub:scenario_1:TMC} - Varying the Number of Devices}
	\label{sub:scenario_1:TMC} 
	
	Here, we study how the number of UEs affects the fairness, energy benefits, and total energy consumption of all UEs. 
	Three ENs with WLAN connections are configured with total resources $(\sum R_j^u, \sum R_j^d, \sum R_j^f)$ = $(108~\textnormal{Mbps},$ $108~\textnormal{Mbps},$ $15~\textnormal{Giga~cycles/s})$ that are enough for $50\% \times 24 = 12$ tasks. Then, we vary the number of UEs $N$ from $2$ to $12$ with different $e_{ij}^u$ and $e_{ij}^d$ increasing by $0.01$~J/Mb from $0.071$~J/Mb to $0.213$~J/Mb. 
	To evaluate the fairness of these schemes, each UE is set an equal demand with $24/N$ tasks. We also set  application compatibility and the highest security level $1$ for all ENs and UEs so that the tasks can be processed locally or at ENs satisfying their requirements. 
	
	
	
	\begin{figure*}[t!]
		\centering
		\begin{subfigure}[t]{0.45\textwidth}
			\centering
			\includegraphics[height=1.6in]{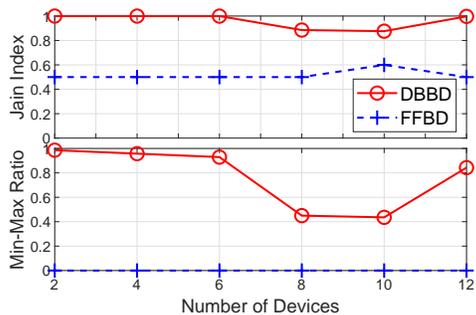}
			\caption{Jain's index and Min-Max Ratio of energy benefits.}
			\label{fig:tmc_scen1_jain_index_min_max_ratio:TMC}
		\end{subfigure}%
		~ 
		\begin{subfigure}[t]{0.45\textwidth}
			\centering
            \includegraphics[height=1.6in]{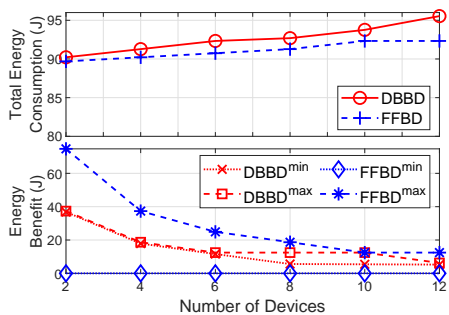}
			\caption{Total consumed energy and energy benefits.}
			\label{fig:tmc_scen1_energy_benefit:TMC}
		\end{subfigure}
		\caption{Jain's index and Min-Max Ratio of energy benefits, total consumed energy, and energy benefits as the number of devices $N$ is increased.}
		\label{fig:tmc_scen1_fairness_indexes_energy_benefit:TMC}
		\vspace{-.2in}
	\end{figure*}

	Figs.~\ref{fig:tmc_scen1_fairness_indexes_energy_benefit:TMC}(a)~and~\ref{fig:tmc_scen1_fairness_indexes_energy_benefit:TMC}(b), respectively, show the fairness indexes and the energy benefits of UEs for the proposed methods when the number of UEs $N$ is increased from $2$ to $12$. From Fig.~\ref{fig:tmc_scen1_fairness_indexes_energy_benefit:TMC}(a), both the Jain's index and min-max ratio in the DBBD are much higher than those in the benchmark, i.e., FFBD. Especially, both indexes are close to their maximum value of $1$ for the cases of $2,~4,~6$,~and~$12$ devices in the DBBD. This is because the DBBD aims to maximize the fairness of energy benefits. Consequently, each UE has an equal number of offloaded tasks, i.e., $6,~3,~2$,~and~$1$, respectively, for these cases. We recall that there is a total of $12$ offloaded tasks as described in the scenario's settings. 
	For the cases of $8$ and $10$ UEs, some have $1$ offloaded task while others have $2$, and consequently, the Jain's index and min-max ratio in these cases are lower than other cases' for the DBBD. 
	As in the scenario's settings, the transmitting/receiving energy consumption units of UE~$n+1$ are $0.01$~J/Mb higher than those of UE~$n$. Thus, for the FFBD that minimizes the total consumed energy, equivalently maximizes the total energy benefits of UEs, $12$ tasks of UEs with less energy consumption are offloaded, whereas $12$ tasks of UEs with higher energy consumption are processed locally. Thus, the Jain's index of the FFBD is mostly close to $0.5$ and the min-max ratio is $0$ for all experiments.
	
	Fig.~\ref{fig:tmc_scen1_fairness_indexes_energy_benefit:TMC}(b) shows that the energy benefits of UEs match the trends of fairness indexes in both methods as in Fig.~\ref{fig:tmc_scen1_fairness_indexes_energy_benefit:TMC}(a). 
	Though the total energy consumption of FFBD is a little lower than that of DBBD, the gap between the minimum ($\textnormal{FFBD}^{\textnormal{min}}$) and maximum ($\textnormal{FFBD}^{\textnormal{max}}$) energy benefits of FFBD is bigger than that of DBBD ($\textnormal{DBBD}^{\textnormal{min}}$ and $\textnormal{DBBD}^{\textnormal{max}}$). This is because the FFBD tries to minimize the total consumed energy, whereas the DBBD aims to maximize the fairness of energy benefits. The zero value of the minimum energy benefits ($\textnormal{FFBD}^{\textnormal{min}}$) also shows that all the tasks of some UEs are processed locally in the FFBD. 

	\subsubsection{Scenario~\ref{sub:scenario_2:TMC} - Varying Edge Nodes' Resources} 
	\label{sub:scenario_2:TMC}
	
	Here, we study how the available resources of ENs affect the fairness, energy benefits, and total energy consumption of all devices. 
	We keep the experiment of $4$ devices in Scenario~\ref{sub:scenario_1:TMC}, in which each device has $6$ tasks. We then vary the total resources of $3$ ENs $(\sum R_j^u,~\sum R_j^d,~\sum R_j^f)$ from $(21.6~\textnormal{Mbps},$ $21.6~\textnormal{Mbps},$ $3~\textnormal{Giga~cycles/s})$  to $(216~\textnormal{Mbps},$ $216~\textnormal{Mbps},$ $30~\textnormal{Giga~cycles/s})$ so that the edge computing can support from $10\%$ to $100\%$ of tasks.
	
	
	
	\begin{figure*}[t!]
		\centering
		\begin{subfigure}[t]{0.45\textwidth}
			\centering
			\includegraphics[height=1.6in]{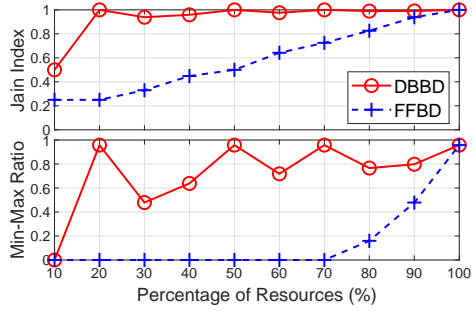}
			\caption{Jain's index and Min-Max Ratio of energy benefits.}
			\label{fig:tmc_scen2_jain_index_min_max_ratio:TMC}
		\end{subfigure}%
		~ 
		\begin{subfigure}[t]{0.45\textwidth}
			\centering
            \includegraphics[height=1.6in]{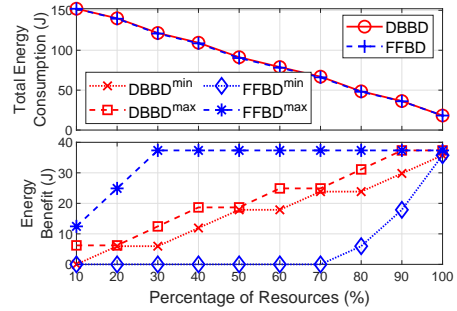}
			\caption{Total consumed energy and energy benefits.}
			\label{fig:tmc_scen2_energy_benefit:TMC}
		\end{subfigure}
		\caption{Jain's index and Min-Max Ratio of energy benefits, total consumed energy, and energy benefits as the ENs' resources are increased.}
		\label{fig:tmc_scen2_fairness_indexes_energy_benefit:TMC}\vspace{-.2in}
	\end{figure*}
	
	Figs.~\ref{fig:tmc_scen2_fairness_indexes_energy_benefit:TMC}(a)~and~\ref{fig:tmc_scen2_fairness_indexes_energy_benefit:TMC}(b), respectively, show the fairness indexes and the energy benefits of UEs for the schemes when the available resources of ENs are increased. From Fig.~\ref{fig:tmc_scen2_fairness_indexes_energy_benefit:TMC}(a), both the Jain's index and min-max ratio in the DBBD are much higher than those in the FFBD. Especially, in the DBBD, the Jain's index is close to $1$ in all experiments except the case the amount of resources is only enough for 10\% of tasks. 
	This is because the more resources the ENs have, the more tasks the UEs can offload. In the DBBD, these offloaded tasks are distributed equally among the UEs to gain the fairness. For example, when the ENs can support processing $60\% \times 24 = 14$ tasks, the $4$ UEs have respectively $4,~4,~3$,~and~$3$ offloaded tasks. As a result, the Jain's index is close to the maximum value of $1$, and the min-max ratio is close to $0.75$.
	As in Scenarios~\ref{sub:scenario_1:TMC}, in the FFBD, the UEs with more energy efficiency from offloading have offloaded tasks, whereas other UEs process theirs tasks locally. Thus, the Jain's index of the FFBD is increased from around $0.25$ to $1$ and its min-max ratio is $0$ for most experiments.
	For example, when the edge computing can support $60\%$, i.e., $14$ tasks, $4$ UEs have, respectively, $6,~6,~2$,~and~$0$ offloaded tasks (i.e., all $6$ tasks of UE~$4$ are processed locally). In this case, the Jain's index is around $0.64$, and the min-max ratio is $0$ when all tasks of UE~$4$ are processed locally.
	
	As in Scenario~\ref{sub:scenario_1:TMC}, Fig.~\ref{fig:tmc_scen2_fairness_indexes_energy_benefit:TMC}(b) shows that the energy consumption of the FFBD is a little lower than that of the DBBD. {\color{black}Here ($\textnormal{DBBD}^{\textnormal{min}}$) and   ($\textnormal{DBBD}^{\textnormal{max}}$) denote the minimum and the maximum energy benefit of all users, respectively. Since we aim to jointly optimize the edge computing, communication resources, and  the task offloading decisions to fairly ``share/allocate'' the energy reduction/benefits to all UEs, the similarity between the two curves ($\textnormal{DBBD}^{\textnormal{min}}$) and   ($\textnormal{DBBD}^{\textnormal{max}}$) mean that the fairness in allocating the energy reduction/benefit to all UEs is achieved. The energy benefit of each UE also matches the trends of fairness indexes in both schemes. Especially, while the gap between the minimum ($\textnormal{DBBD}^{\textnormal{min}}$) and  maximum ($\textnormal{DBBD}^{\textnormal{max}}$) energy benefits of the DBBD is quite small, the gap of the FFBD is very large for most experiments. This is because only the tasks of UEs with less energy consumption are offloaded in the FFBD.} 
	
	\subsubsection{Scenario~\ref{sub:scenario_3:TMC} - Varying Edge Nodes' Resources and Setting the same Devices' Configurations} 
	\label{sub:scenario_3:TMC}
	
	The settings in this scenario are similar to Scenario~\ref{sub:scenario_2:TMC} except for the transmitting/receiving energy consumption units between UEs and ENs are the same as $0.071$~J/Mb. In other words, all devices get the same energy benefits from offloading any computational task. 
	We investigate how three different schemes, i.e., DBBD and FFBD-I/B, return their solutions when the optimal offloading solution may not be unique. 
	
	

 {\color{black}
	
	\begin{figure*}[t!]
		\centering
		\begin{subfigure}[t]{0.45\textwidth}
			\centering
			\includegraphics[height=1.6in]{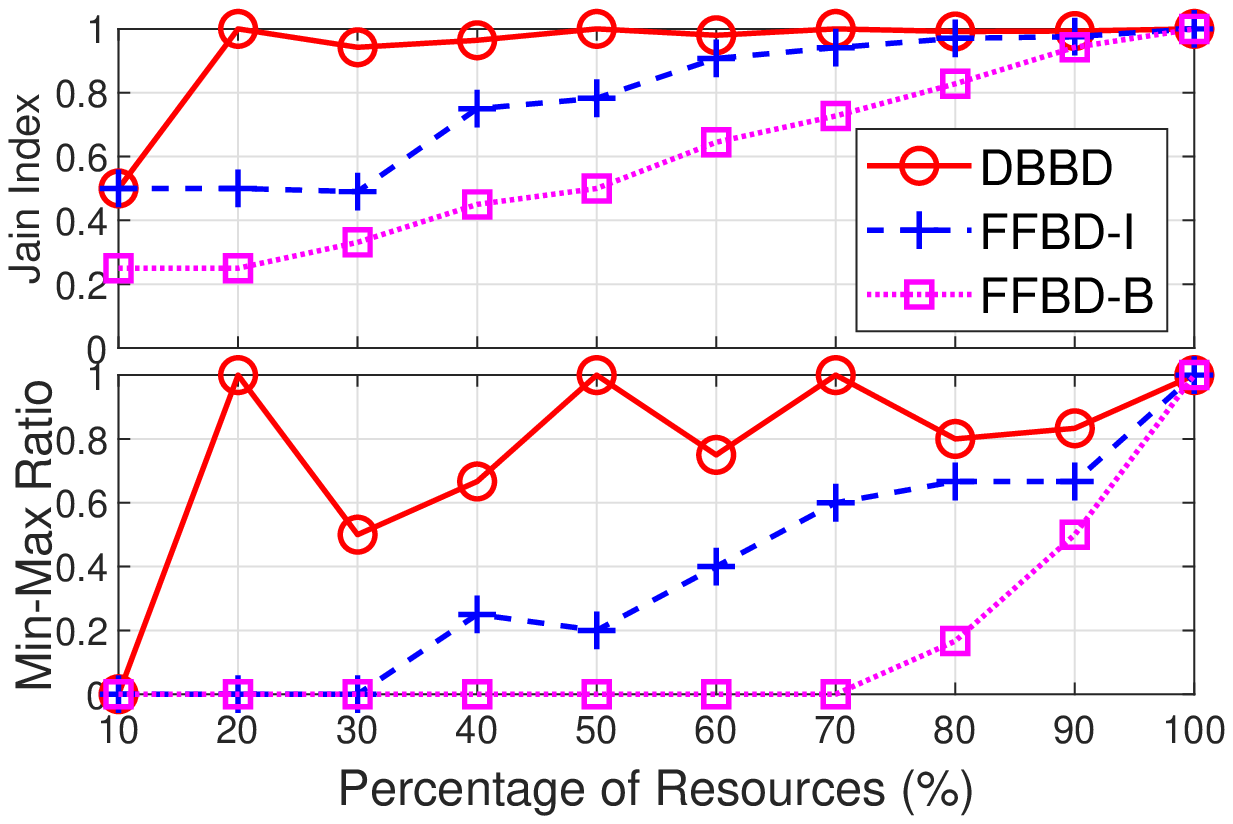}
			\caption{Jain's index and Min-Max Ratio of energy benefits.}
			\label{fig:tmc_scen3_jain_index_min_max_ratio:TMC}
		\end{subfigure}%
		~ 
		\begin{subfigure}[t]{0.45\textwidth}
			\centering
            \includegraphics[height=1.6in]{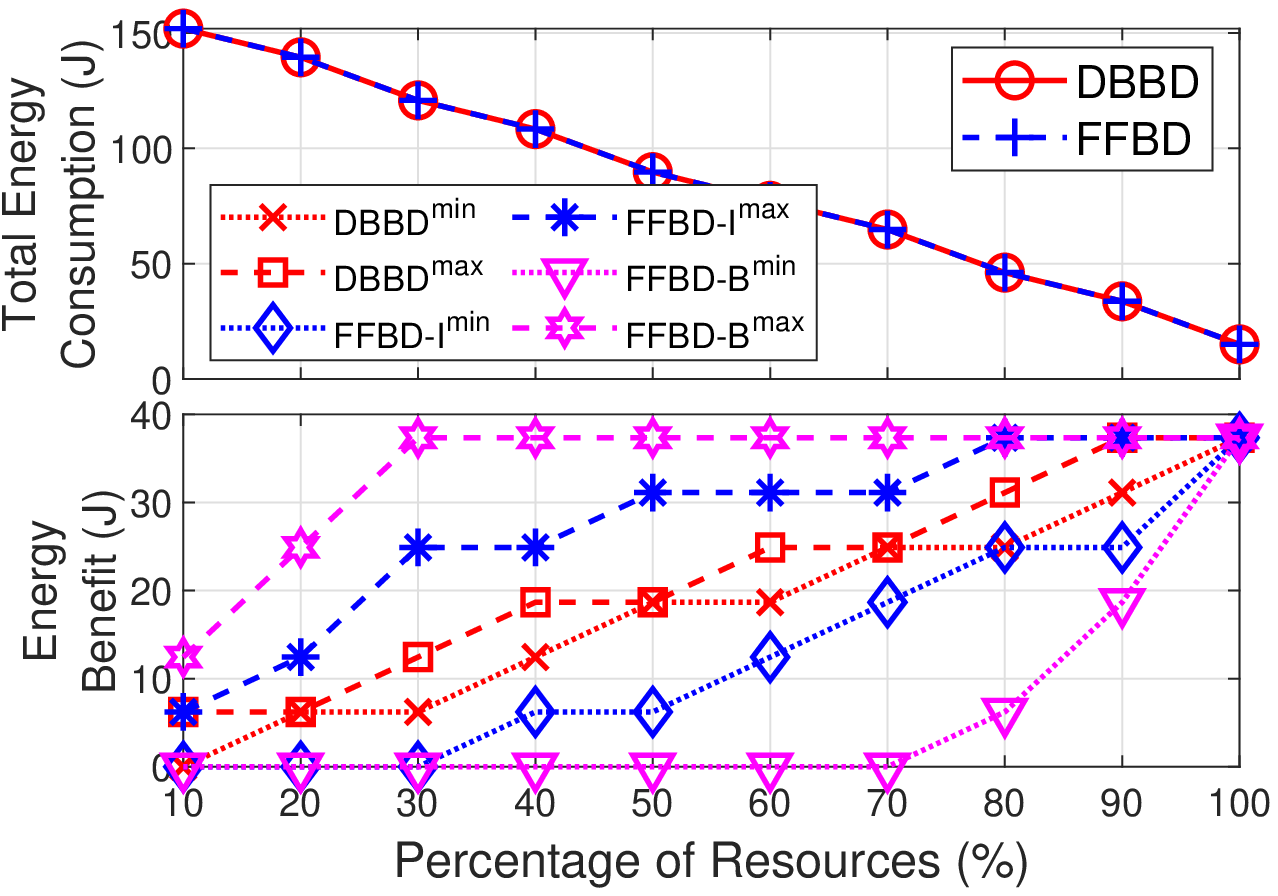}
			\caption{Total consumed energy and energy benefits.}
			\label{fig:tmc_scen3_energy_benefit:TMC}
		\end{subfigure}
		\caption{Jain's index and Min-Max Ratio of energy benefits, total consumed energy, and energy benefits as the ENs' resources are increased while devices' configurations are the same.}
		\label{fig:tmc_scen3_fairness_indexes_energy_benefit:TMC}\vspace{-.2in}
	\end{figure*}}
	
	From Fig.~\ref{fig:tmc_scen3_fairness_indexes_energy_benefit:TMC}(a), the index patterns of the DBBD are similar to those in Scenario~\ref{sub:scenario_2:TMC}. This is because the DBBD maximizes the proportional fairness, which is not affected much by a little difference in the energy benefits of UEs. However, Fig.~\ref{fig:tmc_scen3_fairness_indexes_energy_benefit:TMC}(a) shows that the FFBD-I and FFBD-B return different solutions, though they solve the same problem. The default MOSEK integer solver in the FFBD-I returns any arbitrary offloading solution for the MP, whereas the conventional branch-and-bound algorithm in the FFBD-B tries to offload the tasks of UE~$n$ as much as possible before offloading the tasks of UE~$n+1$. As a result, the Jain's index and min-max ratio of the FFBD-B are the smallest in the three schemes.
	
	In Fig.~\ref{fig:tmc_scen3_fairness_indexes_energy_benefit:TMC}(b), the three schemes, i.e., DBBD and FFBD-I/B, have the same energy consumption, but their maximum and minimum energy benefits between UEs are different.
	Particularly, the gap between the minimum ($\textnormal{FFBD-B}^{\textnormal{min}}$) and maximum ($\textnormal{FFBD-B}^{\textnormal{max}}$) energy benefits of the FFBD-B is bigger than that of the FFBD-I, and the gap of the DBBD (between $\textnormal{DBBD}^{\textnormal{min}}$ and $\textnormal{DBBD}^{\textnormal{max}}$) is the smallest. 
	From this scenario, we can conclude that when an integer problem has multiple optimal solutions, only the DBBD with the proposed DBB for the MP can return the optimal one satisfying the fairness amongst UEs.
	
	\subsubsection{Scenario~\ref{sub:scenario_4:TMC} - Varying the Number of Tasks} 
	\label{sub:scenario_4:TMC}
	
	Here, we study how the number of tasks affects the fairness, energy benefits, and total energy consumption of all UEs. 
	First, two UEs have an equal number of tasks $|\Phi|/2$. We then vary $|\Phi|$ from $2$ to $24$. The UEs have WLAN connections to ENs~$1$~and~$2$ with $e_{ij}^u = e_{ij}^d = 0.071$~J/Mb, and the 3G near connections to EN~$3$. 
	ENs~$1$~and~$2$ have resources $(108~\textnormal{Mbps},$ $108~\textnormal{Mbps},$ $15~\textnormal{Giga~cycles/s})$, the total of that is enough for $24$ tasks. EN~$3$ has resources $(72~\textnormal{Mbps},$ $72~\textnormal{Mbps},$ $10~\textnormal{Giga~cycles/s})$.
	Other parameters are set as in Scenario~\ref{sub:scenario_1:TMC}. 
	
	
	{\color{black}
	\begin{figure*}[t!]
		\centering
		\begin{subfigure}[t]{0.45\textwidth}
			\centering
			\includegraphics[height=1.6in]{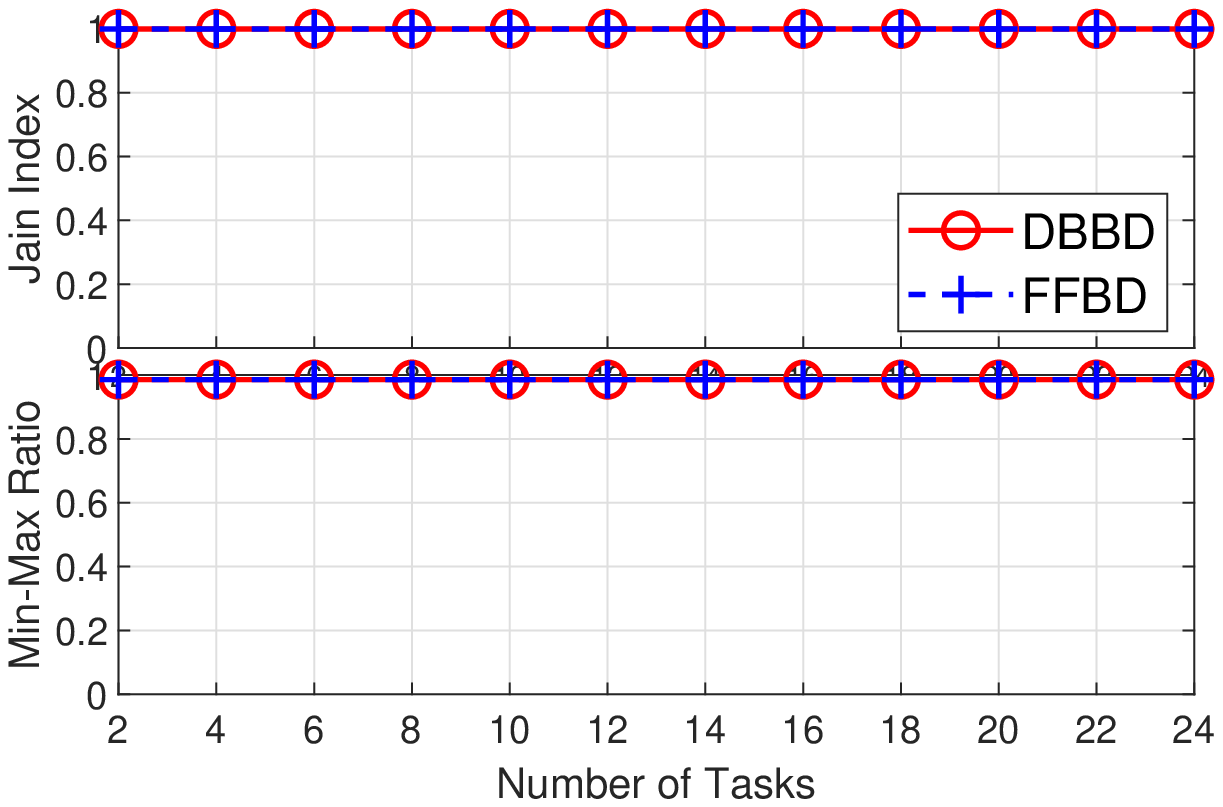}
			\caption{Jain's index and Min-Max Ratio of energy benefits.}
			\label{fig:tmc_scen4_jain_index_min_max_ratio:TMC}
		\end{subfigure}%
		~ 
		\begin{subfigure}[t]{0.45\textwidth}
			\centering
            \includegraphics[height=1.6in]{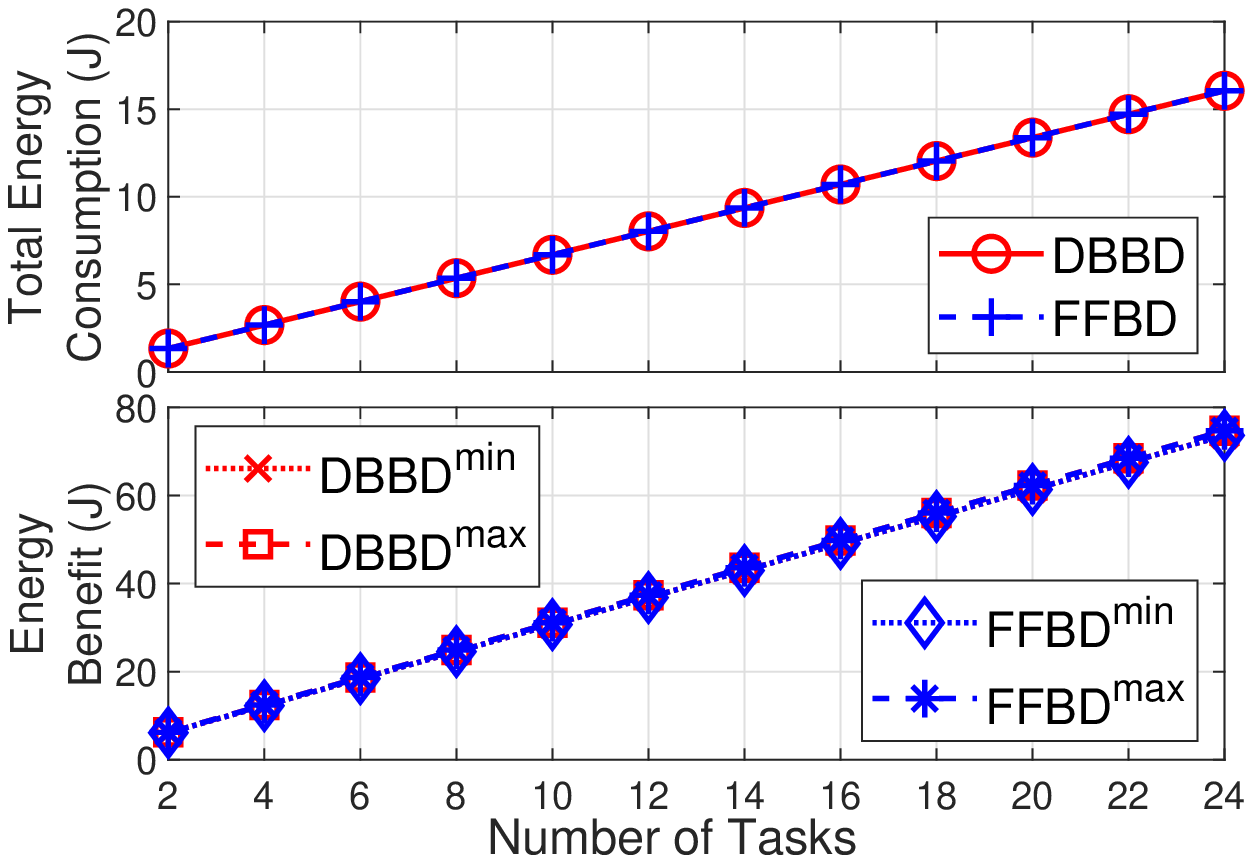}
			\caption{Total consumed energy and energy benefits.}
			\label{fig:tmc_scen4_energy_benefit:TMC}
		\end{subfigure}
		\caption{Jain's index and Min-Max Ratio of energy benefits, total consumed energy, and energy benefits as the number of tasks is increased.}
		\label{fig:tmc_scen4_fairness_indexes_energy_benefit:TMC}
	\end{figure*}}
	
	
	
	\begin{figure*}[t!]
		\centering
		\begin{subfigure}[t]{0.45\textwidth}
			\centering
            \includegraphics[height=1.6in]{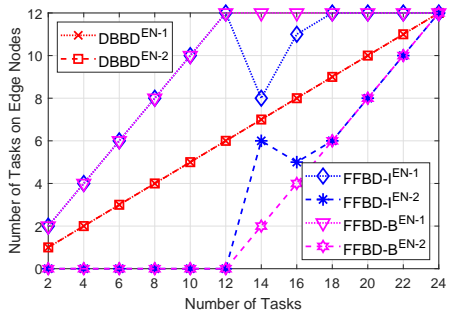}
			\caption{Number of tasks offloaded to each edge node.}
			\label{fig:tmc_scen4_offload_num:TMC}
		\end{subfigure}%
		~ 
		\begin{subfigure}[t]{0.45\textwidth}
			\centering
			\includegraphics[height=1.6in]{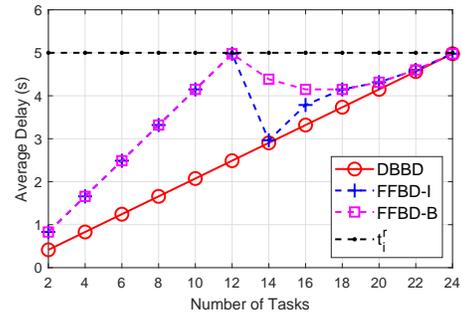}
			\caption{Average delay of tasks.}
			\label{fig:tmc_scen4_delay:TMC}
		\end{subfigure}
		\caption{Number of tasks offloaded to ENs and average delay of tasks as the number of tasks is increased.}\vspace{-.2in}
		\label{fig:tmc_scen4_offload_num_delay:TMC}
	\end{figure*}
	
	From Fig.~\ref{fig:tmc_scen4_fairness_indexes_energy_benefit:TMC}(a), the Jain's index and min-max ratio of both the FFBD and DBBD are approximately equal to $1$ when the number of tasks is increased. Similarly, Fig.~\ref{fig:tmc_scen4_fairness_indexes_energy_benefit:TMC}(b) shows that the DBBD and FFBD have the same total energy consumption and maximum/minimum energy benefits for all experiments. The reason is that  when ENs have sufficient resources to process all the tasks, all the UEs get the maximum energy benefits, whatever scheme (i.e., DBBD or FFBD) is used. These benefits are equal since all UEs have the same configuration and offloading demand.
	
	Figures~\ref{fig:tmc_scen4_offload_num_delay:TMC}(a)~and~\ref{fig:tmc_scen4_offload_num_delay:TMC}(b) show the number of tasks offloaded to each edge node and the average delay of all tasks.
	From Fig.~\ref{fig:tmc_scen4_offload_num_delay:TMC}(a), the DBBD offloads tasks equally to ENs~$1$~and~$2$ (labeled $\textnormal{DBBD}^{\textnormal{EN-1}}$ and $\textnormal{DBBD}^{\textnormal{EN-2}}$). This is  due to the load balance implementation in the DBB algorithm for the MP. The FFBD-I returns an arbitrary offloading decision due to the usage of default solver, whereas the FFBD-B returns the solution in which tasks are offloaded priority to the EN~$1$ and then to EN~$2$. Take the experiment with $16$ tasks as an example, the DBBD, FFBD-I, and FFBD-B, respectively, offload $(8,8)$, $(11,5)$, and $(12,4)$ tasks to ENs-$1$~and~$2$. As a result, the DBBD has a lower average delay than that of the FFBD-I/B have as shown in Fig.~\ref{fig:tmc_scen4_offload_num_delay:TMC}(b). 
	

\subsubsection{Complexity and Computation Time}

 \begin{figure}[h!]
    \centering
    
        \includegraphics[height=1.6in]{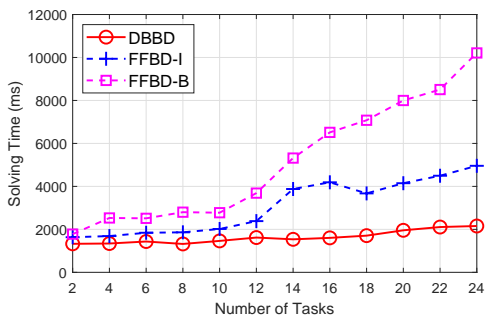}
        
    \caption{{\color{black}Solving time in order to find an optimal solution as the number of tasks is increased.}}
    \label{fig:tmc_solving_time:TMC}
    \end{figure}

{\color{black}

    In this sub-section, we record the solving time of three methods (i.e., DBBD, FFBD-I, and FFBD-B) for different problem sizes by varying the number of tasks (i.e., Scenarios~\ref{sub:scenario_4:TMC}).  Specifically, we vary the number of tasks from $2$ to $24$ and record the solving time, as shown in Fig.~\ref{fig:tmc_solving_time:TMC}.
        Generally, as the number of tasks increases, the computational complexities of all approaches increase.
        This is because all methods require more time to find the solution for the problem with a larger size. Noticeably, the problem size is determined by the number of variables (i.e., offloading variables and resource variables) and constraints (e.g., resource constraints, security, and delay requirements), which are proportional to the number of tasks.
             
        From Fig.~\ref{fig:tmc_solving_time:TMC}, the solving time of all three methods, i.e., DBBD, FFBD-I, and FFBD-B, generally goes up when the problem size is increased. Especially, the proposed DBBD method always has the lowest solving time, and two other methods, i.e., FFBD-I and FFBD-B, respectively, take the second and third positions, irrespective of the number of tasks. This reflects the nature of the three methods. 
        In DBBD, $(\mathbf{P}_0)$ is first decomposed according to integer variables (offloading decisions) and real variables (resource allocations) into a master problem $(\mathbf{MP}_0)$ and subproblems $(\mathbf{SP}_0)$, respectively. The dynamic branch-and-bound algorithm (DBB), which is equipped with an incremental depth-first search, finds the optimal offloading solution of the $(\mathbf{MP}_0)$ considering the load balancing amongst fog/edge nodes. Noticeably, by implementing the incremental depth-first search, the DBB algorithm reuses the search results between iterations, whereas the load balancing returns solutions closer to the optimal one in every iteration. As a result, the solving time of the DBBD method is reduced.
        The FFBD-I uses the default linear \textbf{MOSEK} integer solver to solve the MP, whereas the FFBD-B solves the MP using a conventional branch-and-bound method without the load balancing implementation. In the FFBD-B, tasks are offloaded to EN~$j$ as much as possible before EN~$j+1$. Thus, the FFBD-B method records the highest solving time in all experiments when it solves most of the intermediate problems at nodes on the search tree due to its depth-first search strategy. 
        The MP problem is an integer nonlinear problem, which may have more than one optimal offloading solution. Some offloading solutions can lead to optimal resource allocation solutions for subproblems at fog/edge nodes while others may not. 
        Thus, the solving time of the FFBD-I method slightly fluctuates.
}
	\section{Conclusion} 
	\label{sec:conclusion:TMC}
	{\color{black}
	We have considered the energy-based fairness among user devices in the joint task offloading and resource allocation problem for the multi-layer cooperative edge computing network. To that end, we have formulated a proportional fairness maximization problem that turns out to be NP-hard. To find its optimal solution, we have developed a dynamic branch-and-bound Benders decomposition algorithm (DBBD) to decompose the original problem into subproblems that can be solved parallelly at edge nodes. 
	We have also developed a dynamic branch-and-bound method (DBB), which can solve the master problem with low complexity and satisfy the load balance between edge nodes. 
	We then have compared the DBBD with some benchmarks, namely FFBD-I/B, that optimize the energy consumption without fairness consideration. 
	Numerical results showed that the DBBD always returns the optimal solution, which maximizes the proportional fairness in terms of energy benefits amongst UEs. 
	Using different fairness metrics, i.e., the Jain's index and Min-Max ratio, experiment results also showed that the proposed scheme outperforms the benchmarks, i.e., FFBD-I/B.}
			
	\footnotesize{
\bibliographystyle{IEEEtran}

\bibliography{_DBBD_TMC_REF_Mar2023_submission_recreated_Aug2023,_DBBD_TMC_Additional_Aug2023}

\begin{thebibliography}{10}
\providecommand{\url}[1]{#1}
\csname url@samestyle\endcsname
\providecommand{\newblock}{\relax}
\providecommand{\bibinfo}[2]{#2}
\providecommand{\BIBentrySTDinterwordspacing}{\spaceskip=0pt\relax}
\providecommand{\BIBentryALTinterwordstretchfactor}{4}
\providecommand{\BIBentryALTinterwordspacing}{\spaceskip=\fontdimen2\font plus
\BIBentryALTinterwordstretchfactor\fontdimen3\font minus
  \fontdimen4\font\relax}
\providecommand{\BIBforeignlanguage}[2]{{%
\expandafter\ifx\csname l@#1\endcsname\relax
\typeout{** WARNING: IEEEtran.bst: No hyphenation pattern has been}%
\typeout{** loaded for the language `#1'. Using the pattern for}%
\typeout{** the default language instead.}%
\else
\language=\csname l@#1\endcsname
\fi
#2}}
\providecommand{\BIBdecl}{\relax}
\BIBdecl

\bibitem{vu2022energy}
T.~T. Vu, D.~T. Hoang, K.~T. Phan, D.~N. Nguyen, and E.~Dutkiewicz,
  ``Energy-based proportional fairness for task offloading and resource
  allocation in edge computing,'' in \emph{ICC 2022 - IEEE International
  Conference on Communications}, Conference Proceedings.

\bibitem{Mach2017Mobile}
P.~Mach and Z.~Becvar, ``Mobile edge computing: A survey on architecture and
  computation offloading,'' \emph{IEEE Communications Surveys \& Tutorials},
  vol.~19, no.~3, pp. 1628--1656, 2017.

\bibitem{mao2017survey}
Y.~Mao, C.~You, J.~Zhang, K.~Huang, and K.~B. Letaief, ``A survey on mobile
  edge computing: The communication perspective,'' \emph{IEEE Communications
  Surveys \& Tutorials}, vol.~19, no.~4, pp. 2322--2358, 2017.

\bibitem{el2018edge}
H.~El-Sayed, S.~Sankar, M.~Prasad, D.~Puthal, A.~Gupta, M.~Mohanty, and C.~Lin,
  ``Edge of things: The big picture on the integration of edge, iot and the
  cloud in a distributed computing environment,'' \emph{IEEE Access}, vol.~6,
  pp. 1706--1717, 2018.

\bibitem{du2019enabling}
J.~Du, L.~Zhao, X.~Chu, F.~R. Yu, J.~Feng, and I.~C, ``Enabling low-latency
  applications in lte-a based mixed fog/cloud computing systems,'' \emph{IEEE
  Transactions on Vehicular Technology}, vol.~68, no.~2, pp. 1757--1771, 2019.

\bibitem{xing2019joint}
H.~Xing, L.~Liu, J.~Xu, and A.~Nallanathan, ``Joint task assignment and
  resource allocation for d2d-enabled mobile-edge computing,'' \emph{IEEE
  Transactions on Communications}, vol.~67, no.~6, pp. 4193--4207, 2019.

\bibitem{liu2019dynamic}
C.~Liu, M.~Bennis, M.~Debbah, and H.~V. Poor, ``Dynamic task offloading and
  resource allocation for ultra-reliable low-latency edge computing,''
  \emph{IEEE Transactions on Communications}, vol.~67, no.~6, pp. 4132--4150,
  2019.

\bibitem{tran2019joint}
T.~X. Tran and D.~Pompili, ``Joint task offloading and resource allocation for
  multi-server mobile-edge computing networks,'' \emph{IEEE Transactions on
  Vehicular Technology}, vol.~68, no.~1, pp. 856--868, 2019.

\bibitem{wang2019delay}
J.~Wang, K.~Liu, B.~Li, T.~Liu, R.~Li, and Z.~Han, ``Delay-sensitive
  multi-period computation offloading with reliability guarantees in fog
  networks,'' \emph{IEEE Transactions on Mobile Computing}, pp. 1--1, 2019.

\bibitem{du2018enabling}
J.~Du, L.~Zhao, X.~Chu, F.~R. Yu, J.~Feng, and I.~C.~L, ``Enabling low-latency
  applications in lte-a based mixed fog/cloud computing systems,'' \emph{IEEE
  Transactions on Vehicular Technology}, vol.~68, no.~2, pp. 1757--1771, 2019.

\bibitem{kumar2010cloud}
K.~Kumar and Y.~H. Lu, ``Cloud computing for mobile users: Can offloading
  computation save energy?'' \emph{Computer}, vol.~43, no.~4, pp. 51--56, April
  2010.

\bibitem{vu2021optimal}
T.~T. Vu, D.~N. Nguyen, D.~T. Hoang, E.~Dutkiewicz, and T.~V. Nguyen, ``Optimal
  energy efficiency with delay constraints for multi-layer cooperative fog
  computing networks,'' \emph{IEEE Transactions on Communications}, vol.~69,
  no.~6, pp. 3911--3929, 2021.

\bibitem{wang2019cooperative}
Y.~Wang, X.~Tao, X.~Zhang, P.~Zhang, and Y.~T. Hou, ``Cooperative task
  offloading in three-tier mobile computing networks: An admm framework,''
  \emph{IEEE Transactions on Vehicular Technology}, vol.~68, no.~3, pp.
  2763--2776, 2019.

\bibitem{wang2021fast}
J.~Wang, J.~Hu, G.~Min, A.~Y. Zomaya, and N.~Georgalas, ``Fast adaptive task
  offloading in edge computing based on meta reinforcement learning,''
  \emph{IEEE Transactions on Parallel and Distributed Systems}, vol.~32, no.~1,
  pp. 242--253, 2021.

\bibitem{wang2021dependent}
J.~Wang, J.~Hu, G.~Min, W.~Zhan, A.~Zomaya, and N.~Georgalas, ``Dependent task
  offloading for edge computing based on deep reinforcement learning,''
  \emph{IEEE Transactions on Computers}, pp. 1--1, 2021.

\bibitem{nguyen2014cooperative}
\BIBentryALTinterwordspacing
D.~N. Nguyen and M.~Krunz, ``A cooperative mimo framework for wireless sensor
  networks,'' \emph{ACM Trans. Sen. Netw.}, vol.~10, no.~3, p. Article 43,
  2014. [Online]. Available: \url{https://doi.org/10.1145/2499381}
\BIBentrySTDinterwordspacing

\bibitem{liu2020maxmin}
J.~Liu, K.~Xiong, D.~W.~K. Ng, P.~Fan, Z.~Zhong, and K.~B. Letaief, ``Max-min
  energy balance in wireless-powered hierarchical fog-cloud computing
  networks,'' \emph{IEEE Transactions on Wireless Communications}, vol.~19,
  no.~11, pp. 7064--7080, 2020.

\bibitem{zhang2019femto}
G.~Zhang, F.~Shen, Z.~Liu, Y.~Yang, K.~Wang, and M.~Zhou, ``Femto: Fair and
  energy-minimized task offloading for fog-enabled iot networks,'' \emph{IEEE
  Internet of Things Journal}, vol.~6, no.~3, pp. 4388--4400, 2019.

\bibitem{dong2019energy}
Y.~Dong, S.~Guo, J.~Liu, and Y.~Yang, ``Energy-efficient fair cooperation fog
  computing in mobile edge networks for smart city,'' \emph{IEEE Internet of
  Things Journal}, pp. 1--1, 2019.

\bibitem{li2020auction}
F.~Li, H.~Yao, J.~Du, C.~Jiang, Z.~Han, and Y.~Liu, ``Auction design for edge
  computation ofloading in sdn-based ultra dense networks,'' \emph{IEEE
  Transactions on Mobile Computing}, pp. 1--1, 2020.

\bibitem{liao2020blockchain}
H.~Liao, Y.~Mu, Z.~Zhou, M.~Sun, Z.~Wang, and C.~Pan, ``Blockchain and
  learning-based secure and intelligent task offloading for vehicular fog
  computing,'' \emph{IEEE Transactions on Intelligent Transportation Systems},
  vol.~22, no.~7, pp. 4051--4063, 2021.

\bibitem{zuo2021computation}
Y.~Zuo, S.~Jin, and S.~Zhang, ``Computation offloading in untrusted mec-aided
  mobile blockchain iot systems,'' \emph{IEEE Transactions on Wireless
  Communications}, vol.~20, no.~12, pp. 8333--8347, 2021.

\bibitem{nguyen2018price}
D.~T. Nguyen, L.~B. Le, and V.~Bhargava, ``Price-based resource allocation for
  edge computing: A market equilibrium approach,'' \emph{IEEE Transactions on
  Cloud Computing}, vol.~9, no.~1, pp. 302--317, 2021.

\bibitem{nguyen2019market}
D.~T. Nguyen, L.~B. Le, and V.~K. Bhargava, ``A market-based framework for
  multi-resource allocation in fog computing,'' \emph{IEEE/ACM Transactions on
  Networking}, vol.~27, no.~3, pp. 1151--1164, 2019.

\bibitem{Boyd2004Convex}
S.~Boyd and L.~Vandenberghe, \emph{Convex optimization}.\hskip 1em plus 0.5em
  minus 0.4em\relax Cambridge university press, 2004.

\bibitem{narendra1977branch}
P.~M. Narendra and K.~Fukunaga, ``A branch and bound algorithm for feature
  subset selection,'' \emph{IEEE Transactions on computers}, no.~9, pp.
  917--922, 1977.

\bibitem{jain1984quantitative}
R.~K. Jain, D.-M.~W. Chiu, and W.~R. Hawe, ``A quantitative measure of fairness
  and discrimination,'' \emph{Eastern Research Laboratory, Digital Equipment
  Corporation, Hudson, MA}, 1984.

\bibitem{dang2019trust}
T.~D. Dang, D.~Hoang, and D.~N. Nguyen, ``Trust-based scheduling framework for
  big data processing with mapreduce,'' \emph{IEEE Transactions on Services
  Computing}, vol.~15, no.~1, pp. 279--293, 2022.

\bibitem{razaq2021privacy}
M.~M. Razaq, B.~Tak, L.~Peng, and M.~Guizani, ``Privacy-aware collaborative
  task offloading in fog computing,'' \emph{IEEE Transactions on Computational
  Social Systems}, vol.~9, no.~1, pp. 88--96, 2022.

\bibitem{xiao2021authentication}
H.~Xiao, Q.~Pei, X.~Song, and W.~Shi, ``Authentication security level and
  resource optimization of computation offloading in edge computing systems,''
  \emph{IEEE Internet of Things Journal}, vol.~9, no.~15, pp. 13\,010--13\,023,
  2022.

\bibitem{el2017edge}
H.~El-Sayed, S.~Sankar, M.~Prasad, D.~Puthal, A.~Gupta, M.~Mohanty, and C.~T.
  Lin, ``Edge of things: The big picture on the integration of edge, iot and
  the cloud in a distributed computing environment,'' \emph{IEEE Access},
  vol.~6, pp. 1706--1717, 2018.

\bibitem{lin2015task}
X.~Lin, Y.~Wang, Q.~Xie, and M.~Pedram, ``Task scheduling with dynamic voltage
  and frequency scaling for energy minimization in the mobile cloud computing
  environment,'' \emph{IEEE Transactions on Services Computing}, vol.~8, no.~2,
  pp. 175--186, 2015.

\bibitem{chen2015decentralized}
X.~Chen, ``Decentralized computation offloading game for mobile cloud
  computing,'' \emph{IEEE Transactions on Parallel and Distributed Systems},
  vol.~26, no.~4, pp. 974--983, 2015.

\bibitem{chen2016efficient}
X.~Chen, L.~Jiao, W.~Li, and X.~Fu, ``Efficient multi-user computation
  offloading for mobile-edge cloud computing,'' \emph{IEEE/ACM Transactions on
  Networking}, vol.~24, no.~5, pp. 2795--2808, 2016.

\bibitem{wen2012energy}
Y.~Wen, W.~Zhang, and H.~Luo, ``Energy-optimal mobile application execution:
  Taming resource-poor mobile devices with cloud clones,'' in \emph{2012
  Proceedings IEEE INFOCOM}, Conference Proceedings, pp. 2716--2720.

\bibitem{kelly1998rate}
F.~P. Kelly, A.~K. Maulloo, and D.~K.~H. Tan, ``Rate control for communication
  networks: shadow prices, proportional fairness and stability,'' \emph{Journal
  of the Operational Research Society}, vol.~49, no.~3, pp. 237--252, 1998.

\bibitem{han2005fair}
H.~Zhu, J.~Zhu, and K.~J.~R. Liu, ``Fair multiuser channel allocation for ofdma
  networks using nash bargaining solutions and coalitions,'' \emph{IEEE
  Transactions on Communications}, vol.~53, no.~8, pp. 1366--1376, 2005.

\bibitem{nguyen2015distributed}
D.~N. Nguyen, M.~Krunz, and S.~V. Hanly, ``Distributed bargaining mechanisms
  for mimo dynamic spectrum access systems,'' \emph{IEEE Transactions on
  Cognitive Communications and Networking}, vol.~1, no.~1, pp. 113--127, 2015.

\bibitem{yu2018green}
Y.~Yu, X.~Bu, K.~Yang, and Z.~Han, ``Green fog computing resource allocation
  using joint benders decomposition, dinkelbach algorithm, and modified
  distributed inner convex approximation,'' in \emph{2018 IEEE International
  Conference on Communications (ICC)}, Conference Proceedings, pp. 1--6.

\bibitem{mosek2019documentation}
\BIBentryALTinterwordspacing
E.~D. Andersen and K.~D. Andersen, ``The mosek documentation and api
  reference,'' Report, 2019. [Online]. Available:
  \url{https://www.mosek.com/documentation/}
\BIBentrySTDinterwordspacing

\bibitem{miettinen2010energy}
A.~P. Miettinen and J.~K. Nurminen, ``Energy efficiency of mobile clients in
  cloud computing,'' \emph{HotCloud}, vol.~10, pp. 4--4, 2010.

\bibitem{liu2015small}
F.~Liu, E.~Bala, E.~Erkip, M.~C. Beluri, and R.~Yang, ``Small-cell traffic
  balancing over licensed and unlicensed bands,'' \emph{IEEE Transactions on
  Vehicular Technology}, vol.~64, no.~12, pp. 5850--5865, 2015.

\bibitem{saha2015power}
S.~K. Saha, P.~Deshpande, P.~P. Inamdar, R.~K. Sheshadri, and D.~Koutsonikolas,
  ``Power-throughput tradeoffs of 802.11n/ac in smartphones,'' in \emph{2015
  IEEE Conference on Computer Communications (INFOCOM)}, Conference
  Proceedings, pp. 100--108.

\end{thebibliography}
        }
	
	%
	%
		\begin{IEEEbiography}[{\includegraphics[width=1in,height=1.25in,clip,keepaspectratio]{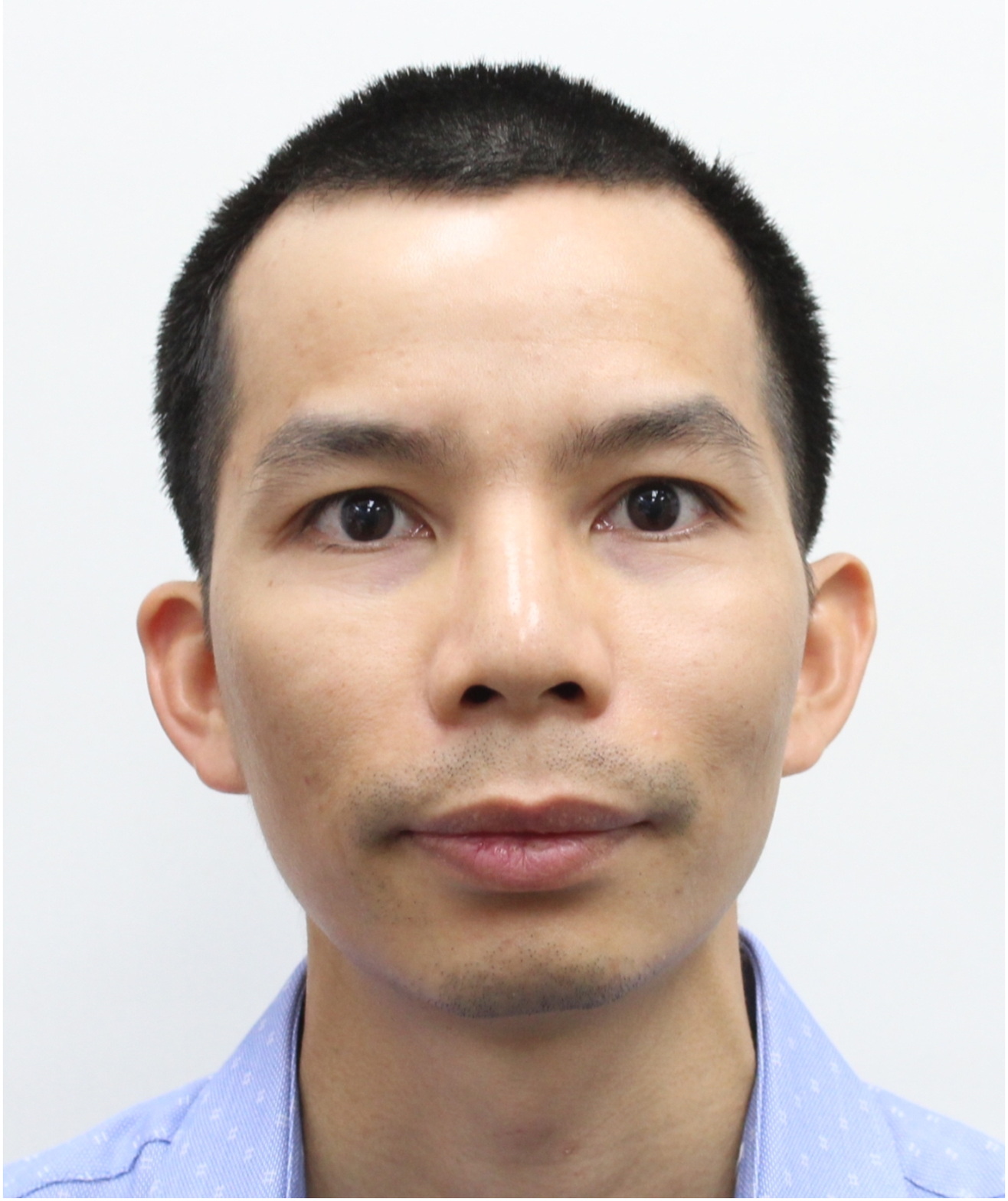}}]{Thai T. Vu} received his B.S. and M.S. degrees in computer science from the VNU University of Engineering and Technology, Hanoi, Vietnam, and his Ph.D. from the University of Technology Sydney (UTS). He is currently a research staff at the School of Engineering and Mathematical Sciences, La Trobe University, Australia. Before pursuing the Ph.D. at UTS, he was a lecturer at the Faculty of Computer Science and Engineering, Thuyloi University, Vietnam. His research interests include fog/cloud computing, Internet of Things, machine learning, and learning algorithms, with an emphasis on energy efficiency, low latency, fairness, and security/privacy awareness.
		\end{IEEEbiography}
		
	\vskip -2\baselineskip plus -1fil
		\begin{IEEEbiography}[{\includegraphics[width=1in,height=1.25in,clip,keepaspectratio]{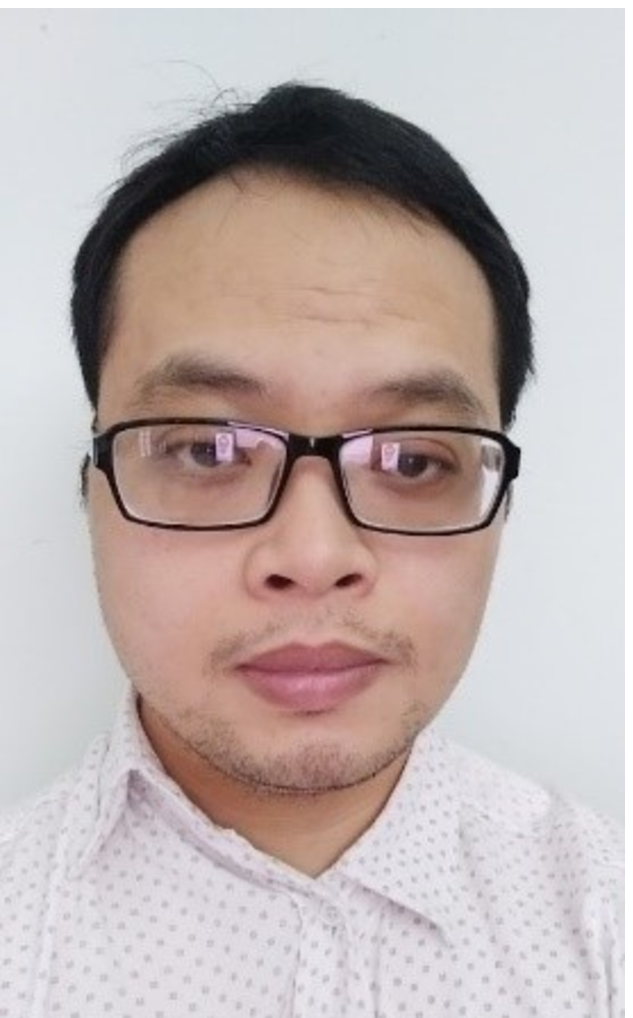}}]{Nam H. Chu} received his B.E. degree in Electronics and Telecommunications Engineering from Hanoi University of Science and Technology, Vietnam in 2009 and his master’s degree in software engineering from the University of Canberra, Australia in 2014. He is currently a Ph.D. student at the University of Technology Sydney, Australia. His research interests include applying machine learning and optimization methods for wireless communication networks.
		\end{IEEEbiography}	
	\vskip -2\baselineskip plus -1fil
		\begin{IEEEbiography}[{\includegraphics[width=1in,height=1.25in,clip,keepaspectratio]{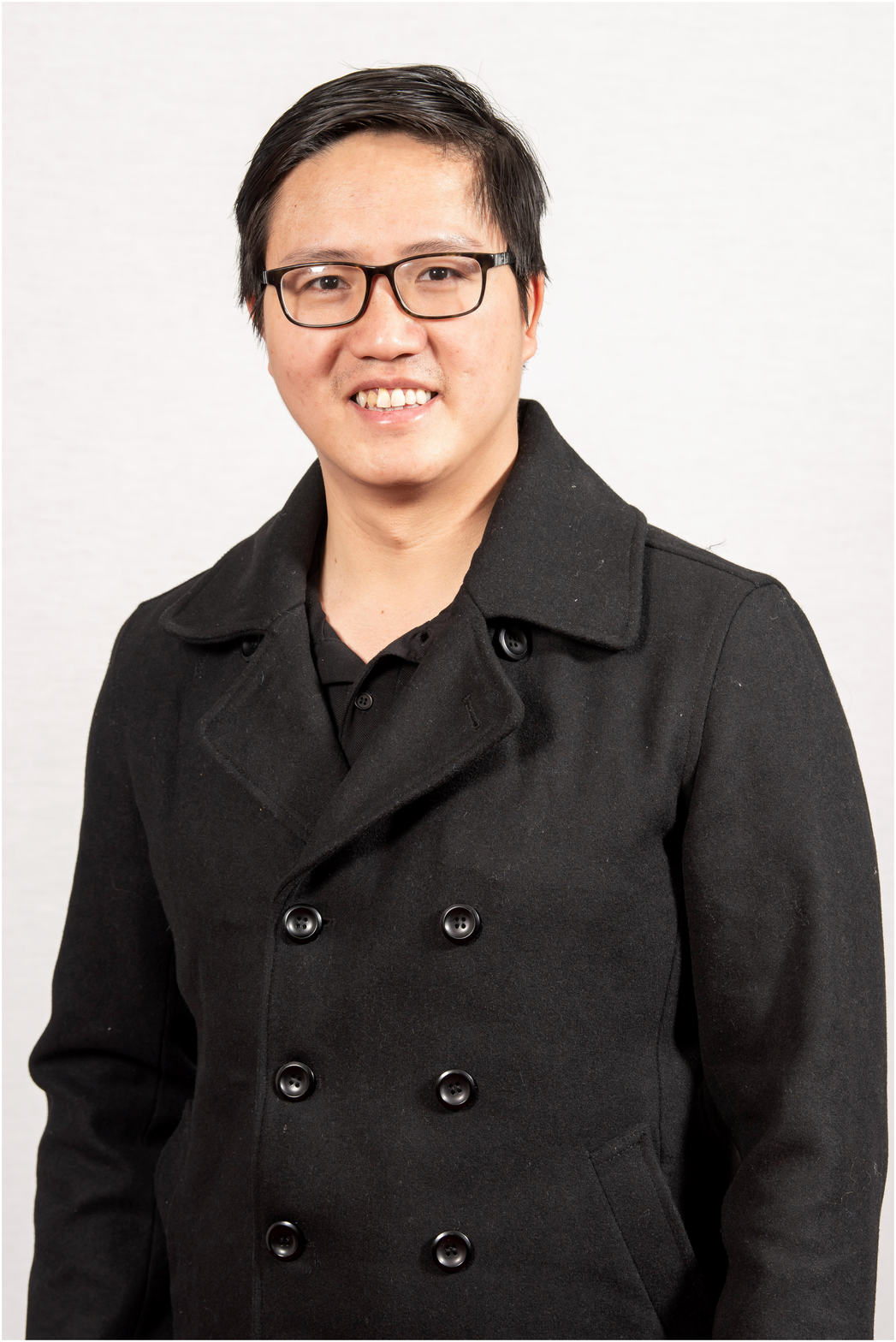}}]{Khoa T. Phan} Phan received the B.Eng. degree in telecommunications (First Class Hons.) from the University of New South Wales (UNSW), Sydney, NSW, Australia, in 2006, the M.Sc. degree in electrical engineering from the University of Alberta, Edmonton, AB, Canada, in 2008, and California Institute of Technology (Caltech), Pasadena, CA, USA, in 2009, respectively, and the Ph.D. degree in electrical engineering from McGill University, Montreal, QC, Canada in 2017.			
		He is currently a Senior Lecturer and Australia Research Council (ARC) Discovery Early Career Researcher Award (DECRA) Fellow with the Department of Computer Science and Information Technology, La Trobe University, Victoria, Australia. His research interests are broadly design, control, optimization, and security of next-generation communications networks with applications in the Internet of Things (IoT), satellite systems, and cloud computing. He is keen on applying machine learning tools such as deep learning, federated learning into designing intelligent secure cyber physical systems. 			
		\end{IEEEbiography}		
		\vskip -2\baselineskip plus -1fil
		\begin{IEEEbiography}[{\includegraphics[width=1in,height=1.25in,clip,keepaspectratio]{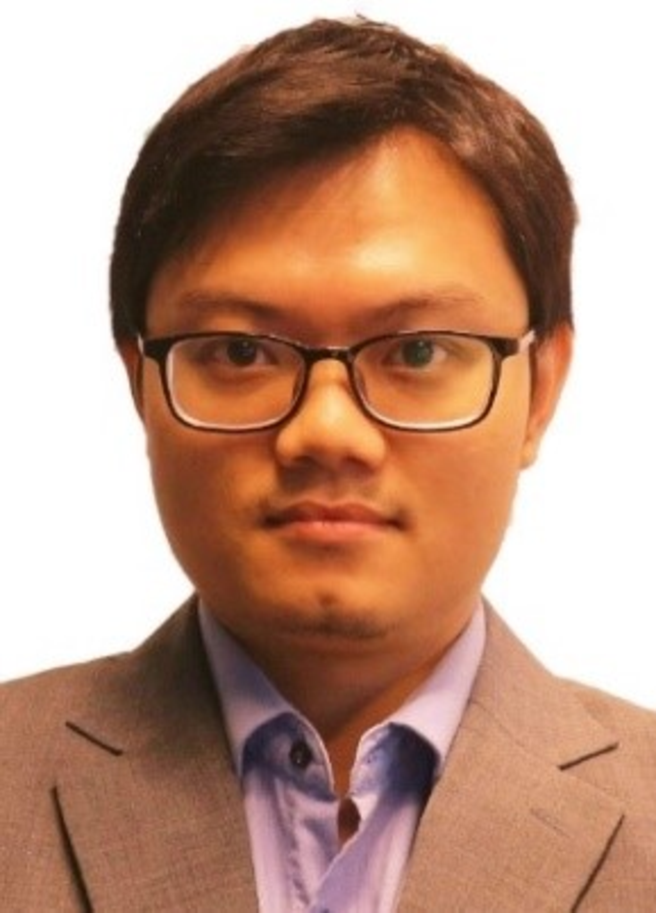}}]{Dinh Thai Hoang}
		(M’16) is currently a faculty member at the School of Electrical and Data Engineering, University of Technology Sydney, Australia. He received his Ph.D. in Computer Science and Engineering from the Nanyang Technological University, Singapore, in 2016. His research interests include emerging topics in wireless communications and networking such as machine learning, ambient backscatter communications, IRS, edge intelligence, cybersecurity, IoT, and 5G/6G networks. He has received several awards including Australian Research Council and IEEE TCSC Award for Excellence in Scalable Computing (Early Career Researcher). Currently, he is an Editor of IEEE Transactions on Wireless Communications, IEEE Transactions on Cognitive Communications and Networking and Associate Editor of IEEE Communications Surveys \& Tutorials.
	\end{IEEEbiography}	
	\vskip -2\baselineskip plus -1fil
		\begin{IEEEbiography}[{\includegraphics[width=1in,height=1.25in,clip,keepaspectratio]{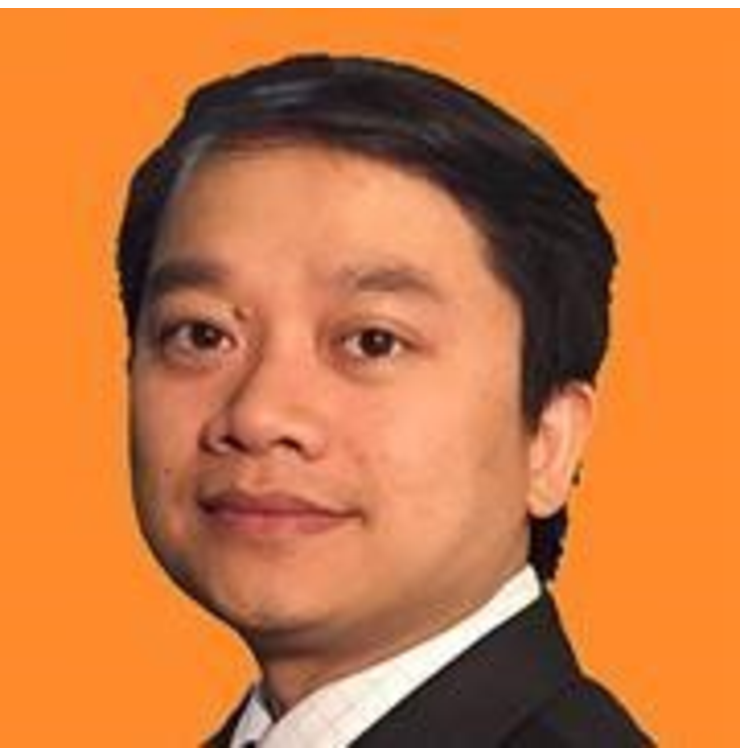}}]{Diep N. Nguyen} (Senior Member, IEEE) received the M.E. degree in electrical and computer engineering from the University of California at San Diego (UCSD) and the Ph.D. degree in electrical and computer engineering from The University of Arizona (UA). He is currently a Faculty Member with the Faculty of Engineering and Information Technology, University of Technology Sydney (UTS). Before joining UTS, he was a DECRA Research Fellow with Macquarie University and a Member of Technical Staff with Broadcom Corporation, Irvine, CA, USA, and ARCON Corporation, Boston, MA, USA, and consulting the Federal Administration of Aviation on turning detection of UAVs and aircraft, and the U.S. Air Force Research Laboratory on anti-jamming. His research interests include computer networking, wireless communications, and machine learning application, with emphasis on systems’ performance and security/privacy. He received several awards from LG Electronics, UCSD, UA, the U.S. National Science Foundation, and the Australian Research Council. He is currently an Editor, Associate Editor of the IEEE Transactions on Mobile Computing and IEEE Open Journal of the Communications Society (OJ-COMS).
		\end{IEEEbiography}	
	\vskip -2\baselineskip plus -1fil
		\begin{IEEEbiography}[{\includegraphics[width=1in,height=1.25in,clip,keepaspectratio]{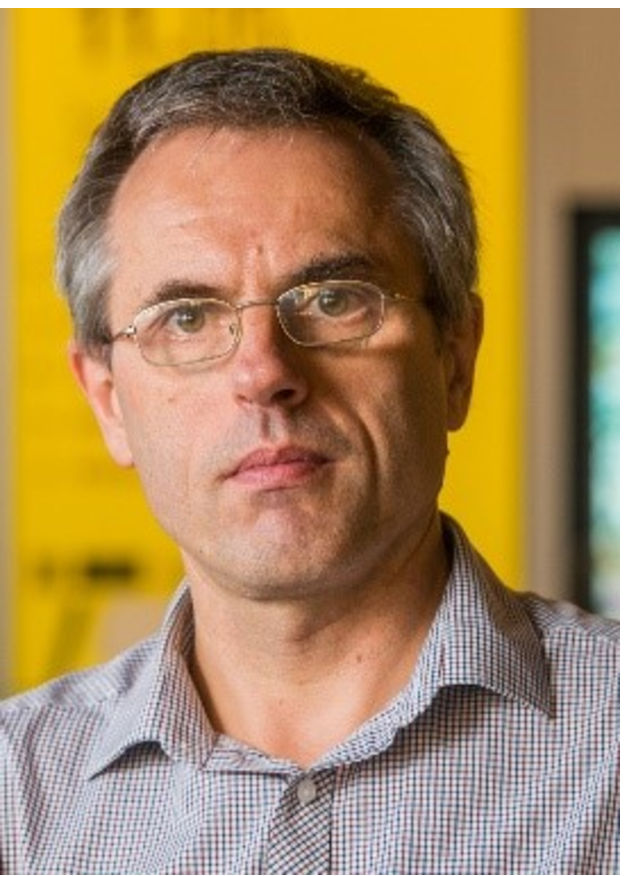}}]{Eryk Dutkiewicz} received his B.E. degree in Electrical and Electronic Engineering from the University of Adelaide in 1988, his M.Sc. degree in Applied Mathematics from the University of Adelaide in 1992 and his PhD in Telecommunications from the University of Wollongong in 1996. His industry experience includes management of the  Wireless Research Laboratory at Motorola in early 2000’s. Prof. Dutkiewicz is currently the Head of School of Electrical and Data Engineering at the University of Technology Sydney, Australia. He is a Senior Member of IEEE. He also holds a professorial appointment at Hokkaido University in Japan. His current research interests cover 5G/6G and IoT networks.
		\end{IEEEbiography}
	\vfill 
	\newpage
	\appendices
	\section{Proof of Theorem~\ref{theo:convexity:TMC}}
	\label{sec:theo_convexity:TMC}
	
	\begin{proof} First, we will show that the objective~$\sum_{n=1}^{N} \rho_n \text{ln}(u_n)$ is concave w.r.t. $\left(\mathbf{x}, \mathbf{r}, \mathbf{b}\right)$. Obviously, $\text{ln}(u_n)$ is concave w.r.t. $(u_n)$.  From Eqs.~(\ref{eq:local_en:TMC}),~(\ref{eq:edge_en:TMC}),~(\ref{eq:indirect_cloud_en:TMC}),~(\ref{eq:direct_cloud_en:TMC}),~(\ref{eq:task_en:TMC}),~and~(\ref{eq:utility_fuc_:TMC}), the utility $u_n$ is a linear function of variables $\mathbf{x}$ because $(E_{i}^{base} - \mathbf{e}_i)$ (i.e., in $u_n = \sum_{I_i \in \Phi_n}\Delta_{i} = \sum_{I_i \in \Phi_n}(E_{i}^{base} - \mathbf{e}_i)^{\top}\mathbf{x}_{i} $) is a constant vector. Thus, $\text{ln}(u_n)$ is a concave function w.r.t.  $\left(\mathbf{x}, \mathbf{r}, \mathbf{b}\right)$ according to the rule of composition with an affine mapping~\cite{Boyd2004Convex}. Due to $\rho_n > 0, \forall 1 \leq n \leq N$, we have $\sum_{n=1}^{N} \rho_n \text{ln}(u_n)$ is a concave function w.r.t.  $\left(\mathbf{x}, \mathbf{r}, \mathbf{b}\right)$.

		Second, we will show that all constraint functions in $(\mathbf{R}_0)$ and $(\mathbf{\widetilde{X}}_0)$ are convex. 
		From  Eqs.~(\ref{eq:local_delay:TMC}),~(\ref{eq:edge_delay:TMC}),~(\ref{eq:indirect_cloud_delay:TMC}), (\ref{eq:direct_cloud_delay:TMC}), and~(\ref{eq:task_delay_convex:TMC}), the delay $T_{i}=\mathbf{h}_i^{\top}\mathbf{y}_{i}$ is the sum of functions, i.e., $x_{i}^{l2}$, $\frac{x_{ij}^{f2}}{r_{ij}^{u}}$, $\frac{x_{ij}^{f2}}{r_{ij}^{d}}$, $\frac{x_{ij}^{f2}}{r_{ij}^{f}}$, $x_{ij}^{c2}$, $\frac{x_{ij}^{c2}}{r_{ij}^{u}}$, $\frac{x_{ij}^{c2}}{r_{ij}^{d}}$, and $\frac{x_{ij}^{c2}}{b_{ij}}$ $\forall j \in \mathcal{M}$, with positive coefficients, e.g., $\frac{L_{i}^{u} w_{i}}{f_{i}^{l}}$, $L_i^u$, $L_i^d$, $(L_{i}^{u} w_{i})$, and $\left(L_{i}^{u}+L_{i}^{d}\right)$. The function $x^2$ is convex. We need to prove that function $g(x,r) = \frac{x^2}{r}$ is convex. 
		Let $\mathbf{H} = \nabla^2 g(x,r)$ be the Hessian of $g(x,r)$.
		\begin{equation}
			\mathbf{H} = 	\left[	\begin{array}{cc}
				\frac{\partial^2 g}{\partial^2 x} & \frac{\partial^2 g}{\partial x \partial r} \\
				\frac{\partial^2 g}{\partial r \partial x} & \frac{\partial^2 g}{\partial^2 r}	
			\end{array}	\right]
			= \left[	\begin{array}{cc}
				\frac{2}{r} & -\frac{2x}{r^2} \\
				-\frac{2x}{r^2} & \frac{2x^2}{r^3}	
			\end{array}	\right].
			\label{eq:Hessian_matrix:TMC}
		\end{equation}
		
		Then, given an arbitrary real vector $\mathbf{v} = (v_1,v_2)$, we have
		\begin{equation}
			\begin{split}
				\mathbf{v}^\top \mathbf{H} \mathbf{v} &= v_1 \left(v_1 \frac{2}{r} - v_2 \frac{2x}{r^2} \right) + v_2 \left(-v_1 \frac{2x}{r^2} + v_2 \frac{2x^2}{r^3} \right) \\
				&= \frac{2}{r} \left(v_1 - v_2 \frac{x}{r} \right)^2.
			\end{split}
			\label{eq:Hessian_matrix_quadratic:TMC}
		\end{equation}
		
		The variable $r$ in Eq.~(\ref{eq:Hessian_matrix_quadratic:TMC}) is a representation of the variables $r_{ij}^u, r_{ij}^d, r_{ij}^f,b_{ij} \geq 0$. Thus, we have $r \geq 0$ and  $\mathbf{v}^\top \mathbf{H} \mathbf{v} \geq 0$. In other words, $\mathbf{H}$ is positive semidefinite, and thus $g(x,r)$ is convex w.r.t. $(x,r)$~\cite{Boyd2004Convex}. As a result, $T_{i}$ is convex since it is the nonnegative weighted sum of convex functions. Particularly, $(\mathcal{C}_1)$ in $(\mathbf{R}_0)$ is a convex function w.r.t. $\left(\mathbf{x}, \mathbf{r},\mathbf{b}\right)$. Besides, $(\mathcal{C}_{2-9})$ in $(\mathbf{R}_0)$ and $(\mathbf{\widetilde{X}}_0)$ are linear functions. 
		
		Since the relaxed problem $(\mathbf{\widetilde{P}}_0)$ aims to maximize the concave objective over the feasible convex set defined by $(\mathbf{R}_0)$ and $(\mathbf{\widetilde{X}}_0)$, the   $(\mathbf{\widetilde{P}}_0)$ is a convex optimization problem~\cite{Boyd2004Convex}.
	\end{proof}

	\section{Proof of Theorem~\ref{theo:dbbdstop:TMC}}
	\label{sec:theo_dbbdstop:TMC}
	
	
	\begin{proof}
		We have $cuts^{(k)}$ and $cuts^{(k+1)}$, respectively, are the sets of Benders cuts of $(\mathbf{MP}_0)$ at iterations $(k)$ and $(k+1)$.
		At iteration $k$, we assume that $(\mathbf{MP}_0)$ has a solution $\mathbf{x}^{(k)}$, which leads to at least one infeasible subproblem~$(\mathbf{SP}_1)$.
		Consequently, a new subproblem Benders cut will be added to $cuts^{(k+1)}$, thus we have $cuts^{(k)} \subset cuts^{(k+1)}$.
		This leads to $\underset{\mathbf{x} \in \mathbf{X}_0}{\max} \{\sum_{n=1}^{N} \rho_n \text{ln}(u_n) \}$ s.t. $cuts^{(k)}$ $\geq$ $\underset{\mathbf{x} \in \mathbf{X}_0}{\max} \{\sum_{n=1}^{N} \rho_n \text{ln}(u_n) \}$ s.t. $cuts^{(k+1)}$. In other words, $\underset{\mathbf{x} \in \mathbf{X}_0}{\max} \{\sum_{n=1}^{N} \rho_n \text{ln}(u_n) \}$ s.t. $cuts^{(k)}$ is a decreasing function with iteration $k$.
		Thus, the first found solution $(\mathbf{x},\mathbf{r},\mathbf{b})$ of $(\mathbf{MP}_0)$ and $(\mathbf{SP}_0)$ is the optimal solution of~$(\mathbf{P}_0)$.
		If $(\mathbf{MP}_0)$ has no solution at iteration $k$, it will not have any solution at subsequent iterations because $cuts^{(k)} \subset cuts^{(k+v)}, \forall v \geq 1$. Hence, we can conclude the unfeasibility of the original problem $(\mathbf{P}_0)$.
	\end{proof}

	\section{Proof of Theorem~\ref{theo:feasible:TMC}}
	\label{sec:theo_feasible:TMC}
	\begin{proof}
		With the resource allocation $\mathbf{r}_{ij}=(r_{ij}^{u},r_{ij}^{d},r_{ij}^{f}, b_{ij}) =  (\frac{L_i^{u'}}{\beta_{bal}^u},  \frac{L_i^{d'}}{\beta_{bal}^d}, \frac{L_i^{u'}w_i^{'}}{\beta_{bal}^f}, \frac{L_i^{c'}}{\beta_{bal}^b})$ toward task $I_i$,  
		we have $\beta_i =
		\frac{L_i^{u'}}{r_{ij}^u}+ \frac{L_i^{d'}}{r_{ij}^d}+
		\frac{L_i^{u'}w_i^{'}}{r_{ij}^f} + \frac{L_i^{c'}}{b_{ij}} = 
		(\beta_{bal}^u + \beta_{bal}^d + \beta_{bal}^f  + \beta_{bal}^b) = \beta_{bal},  \forall i \in \Phi_j^{t+s}$. 
		Here, $r_{ij}^{f} = 0$ and $\frac{L_i^{u'}w_i^{'}}{r_{ij}^{f}} = 0, \forall i \in \Phi_j^s$, whereas $b_{ij} = 0$ and $\frac{L_i^{c'}}{b_{ij}} = 0, \forall i \in \Phi_j^t$. 
		Thus, $\beta_i = \beta_{bal} \leq 1, \forall i \in \Phi_j^{t+s}$.
		
		Besides, $\sum_{i \in \Phi_j^{t+s}}r_{ij}^u = R_j^u$, $\sum_{i \in \Phi_j^{t+s}}r_{ij}^d = R_j^d$, $\sum_{i \in \Phi_j^{t+s}}r_{ij}^f = R_j^f$, and $\sum_{i \in \Phi_j^{t+s}}b_{ij} = \mathcal{B}_j$ satisfying resource constraints at EN~$j$. To conclude, $(\mathbf{SP}_1)$ is feasible with the  solution $\mathbf{r}_{ij}$.	
	\end{proof}

	%

	\section{Proof of Theorem~\ref{theo:infeasible:TMC}}
	\label{sec:theo_infeasible:TMC}
	\begin{proof}
		
		We first prove the following Lemma~\ref{lem:inequality:TMC} to support Theorem~\ref{theo:infeasible:TMC}.
		\begin{lemma}
			\label{lem:inequality:TMC}
			
			Given two sequences of numbers $p_i \geq 0$, $q_i > 0$, $\forall i \in N$. 
			We have $\underset{i \in N}{\max} \{\frac{p_i}{q_i}\} \geq \frac{\sum_{i \in N} p_i}{\sum_{i \in N} q_i}$.
		\end{lemma}
		
		\begin{proof}
			If $\frac{p_1}{q_1} \geq \frac{p_2}{q_2}$ then $\max\{\frac{p_1}{q_1}, \frac{p_2}{q_2}\} = \frac{p_1}{q_1} \geq \frac{p_1+p_2}{q_1+q_2}$. Otherwise, if $\frac{p_1}{q_1} < \frac{p_2}{q_2}$ then $\max\{\frac{p_1}{q_1}, \frac{p_2}{q_2}\} = \frac{p_2}{q_2} > \frac{p_1+p_2}{q_1+q_2}$. Thus, $\max\{\frac{p_1}{q_1}, \frac{p_2}{q_2}\} \geq \frac{p_1+p_2}{q_1+q_2}$. 
			Similarly, $\max\{\frac{p_1+p_2}{q_1+q_2}, \frac{p_3}{q_3}\} \geq \frac{p_1+p_2+p_3}{q_1+q_2+q_3}$. Therefore, $\max\{\frac{p_1}{q_1}, \frac{p_2}{q_2}, \frac{p_3}{q_3}\} \geq \max\{\frac{p_1+p_2}{q_1+q_2}, \frac{p_3}{q_3}\} \geq \frac{p_1+p_2+p_3}{q_1+q_2+q_3}$. 
			Repeatedly, we have $\underset{i \in N}{\max} \{\frac{p_i}{q_i}\} \geq \frac{\sum_{i \in N} p_i}{\sum_{i \in N} q_i}$.
		\end{proof}

		Applying Lemma~\ref{lem:inequality:TMC} into $\{L_i^{u'}\}_{i \in \Phi_j^{t+s}}$ and $\{r_{ij}^{u}\}_{i \in \Phi_j^{t+s}}$, we have $\underset{i \in \Phi_j^{t+s}}{\max} \{\frac{L_i^{u'}}{r_{ij}^{u}}\} \geq \frac{\sum_{i \in \Phi_j^{t+s}}{L_i^{u'}}}{\sum_{i \in \Phi_j^{t+s}}r_{ij}^{u}}$. 	
		According to resource allocation conditions, $\sum_{i \in \Phi_j^{t+s}}r_{ij}^{u} \leq R_j^u$, we have $\underset{i \in \Phi_j^{t+s}}{\max} \{\frac{L_i^{u'}}{r_{ij}^{u}}\} \geq \frac{\sum_{i \in \Phi_j^{t+s}}{L_i^{u'}}}{R_j^u}$. Therefore, $\underset{i \in \Phi_j^{t+s}}{\max} \{\frac{L_i^{u'}}{r_{ij}^{u}}\} > 1$. Without loss of generality, we assume $\exists i^* \in \Phi_j^{t+s}, \frac{L_{i^*}^{u'}}{r_{i^*j}^{u}} = \underset{i \in \Phi_j^{t+s}}{\max} \{\frac{L_i^{u'}}{r_{ij}^{u}}\} > 1$. Consequently, $\beta_{i^*} \!=\!(
		\frac{L_{i^*}^{u'}}{r_{i^*j}^u}+$ 
		$\frac{L_{i^*}^{d'}}{r_{i^*j}^d}+$ 
		$\frac{L_{i^*}^{u'}w_i^{'}}{r_{i^*j}^f} +$  $\frac{L_{i^*}^{c'}}{b_{i^*j}}) \!>\!$  $\frac{L_{i^*}^{u'}}{r_{i^*j}^{u}} \!>\! 1$.
		It contradicts the delay requirement of Task $I_{i^*}$, $\beta_{i^*} \leq 1$ as in  Eq.~(\ref{eq:delay_constraints_new2:TMC}). Thus, the problem $(\mathbf{SP}_1)$ is infeasible.
		
		Other cases, i.e., $\frac{\sum_{i \in \Phi_j^{t+s}}
			L_i^{d'}}{R_j^d} > 1$,  $\frac{\sum_{i \in \Phi_j^{t+s}}
			L_i^{u'}w_i^{'}}{R_j^f} > 1$, and $\frac{\sum_{i \in \Phi_j^{t+s}}
			L_i^{c'}}{\mathcal{B}_j} > 1$, are proved in a similar way.
	\end{proof}

	\section{Proof of Theorem~\ref{theo:convexity_sub:TMC}}
	\label{sec:theo_convexity_sub:TMC}
	
	\begin{proof} The objective~$\gamma_{j}$ is linear w.r.t. $\left(\mathbf{r}_j, \mathbf{b}_j, \gamma_{j}\right)$. 
		We need to show that all constraint functions in $(\mathbf{\widetilde{R}}_j)$ are convex. 
		From Eq.~(\ref{eq:delay_constraints_new2:TMC})~and~$(\mathcal{C}_{1j})$ in Eq.~(\ref{eq:resource_and_delay_con:TMC}), the delay constraint $\beta_i \leq \gamma_{j}$ can be rewritten as 
		$(
		\frac{L_{i}^{u'}}{r_{ij}^{u}}+
		\frac{L_{i}^{d'}}{r_{ij}^{d}}+
		\frac{L_{i}^{u'}w_{i}^{'}}{r_{ij}^{f}}+
		\frac{L_{i}^{c'}}{b_{ij}}
		) - \gamma_{j} \leq 0$. 
		This function is the sum of convex functions, i.e., $\frac{1}{r_{ij}^{u}}$, $\frac{1}{r_{ij}^{d}}$, $\frac{1}{r_{ij}^{f}}$, $\frac{1}{b_{ij}}$, and $-\gamma_{j}$ with positive coefficients, e.g., $L_{i}^{u'}$, $L_{i}^{d'}$, $L_{i}^{u'}w_{i}^{'}$, $L_{i}^{c'}$, and $1$. 
		Thus, the function $(
		\frac{L_{i}^{u'}}{r_{ij}^{u}}+
		\frac{L_{i}^{d'}}{r_{ij}^{d}}+
		\frac{L_{i}^{u'}w_{i}^{'}}{r_{ij}^{f}}+
		\frac{L_{i}^{c'}}{b_{ij}}
		) - \gamma_{j}$ is convex w.r.t. $\left(\mathbf{r}_j, \mathbf{b}_j, \gamma_{j}\right)$. 
		Besides, all other constraints in Eq.~(\ref{eq:resource_and_delay_con:TMC}) are linear functions. 		
		Since the subproblem $(\mathbf{SP}_2)$ aims to minimize the convex objective function over the feasible convex set defined by $(\mathbf{\widetilde{R}}_j)$, the subproblem $(\mathbf{SP}_2)$ is a convex optimization problem~\cite{Boyd2004Convex}.
	\end{proof}

\end{document}